\documentclass[aoas]{imsart}

\RequirePackage{amsthm,amsmath,amsfonts,amssymb}
\RequirePackage{bbm}
\RequirePackage[authoryear]{natbib}
\RequirePackage[colorlinks,citecolor=blue,urlcolor=blue]{hyperref}
\RequirePackage{graphicx}
\RequirePackage{scrextend}
\RequirePackage[ruled,norelsize,vlined]{algorithm2e}
\RequirePackage{colortbl} 
\RequirePackage{xcolor}
\RequirePackage{booktabs}
\usepackage[T1]{fontenc}
\usepackage{tgpagella}

\changefontsizes{13.3pt}

\startlocaldefs

\newcommand{\bY}{ {\bf Y} }
\newcommand{\by}{ {\bf y} }
\newcommand{\bz}{ {\bf z} }
\newcommand{\bTheta}{ {\boldsymbol \Theta} }
\newcommand{\bpi}{ {\boldsymbol \pi} }
\newcommand{\bbeta}{ {\boldsymbol \beta} }
\newcommand{\bX}{ {\bf X} }
\newcommand{\bnu}{ {\boldsymbol \nu} }
\newcommand{\balpha}{ {\boldsymbol \alpha} }

\theoremstyle{plain}

\theoremstyle{remark}

\newtheorem*{example}{Example}

\endlocaldefs

\begin{document}

\begin{frontmatter}
\title{Extended stochastic block models \\ with application to criminal networks}
\runtitle{Extended stochastic block models}

\begin{aug}
\author[A]{\fnms{Sirio} \snm{Legramanti}\ead[label=e1,mark]{sirio.legramanti@unibocconi.it}},
\author[B]{\fnms{Tommaso} \snm{Rigon}\ead[label=e2,mark]{tommaso.rigon@unimib.it}}
\\
\author[A]{\fnms{Daniele} \snm{Durante}\ead[label=e3,mark]{daniele.durante@unibocconi.it}}
\and
\author[C]{\fnms{David B.} \snm{Dunson}\ead[label=e4,mark]{dunson@duke.edu}}
\address[A]{Department Decision Sciences and Institute for Data Science and Analytics,
Bocconi University,
\printead{e1,e3}}

\vspace{-5pt}

\address[B]{Department of Economics, Management and Statistics,
University of Milano-Bicocca,
\printead{e2}}

\vspace{-5pt}

\address[C]{Department of Statistical Science,
Duke University,
\printead{e4}}
\end{aug}

\begin{abstract}
Reliably learning group structures among nodes in network data is challenging in several applications.  We are particularly motivated by studying covert networks that encode relationships among  criminals. These  data are subject to measurement errors,  and  exhibit a complex combination  of an unknown number of core-periphery, assortative and disassortative  structures that may unveil key architectures of the criminal organization.  The coexistence of these noisy block patterns limits the reliability of routinely-used community detection algorithms, and requires extensions of model-based solutions to realistically characterize the node partition process, incorporate information from node attributes, and provide improved strategies for estimation and uncertainty quantification. To cover these gaps, we develop a new class of extended stochastic block models (\textsc{esbm}) that infer groups of nodes having common connectivity patterns via Gibbs-type priors on the partition process. This choice encompasses many realistic priors for criminal networks, covering solutions with fixed, random and infinite number of possible groups, and facilitates the inclusion of  node attributes in a principled manner. Among the new alternatives in our class, we focus on the Gnedin process as a realistic prior that allows the number of groups to be finite, random and subject to a reinforcement process coherent with criminal networks. A  collapsed Gibbs sampler is proposed for the whole \textsc{esbm} class, and refined strategies for estimation, prediction, uncertainty quantification and model selection are outlined. The  \textsc{esbm} performance is illustrated in realistic simulations and in an application to an Italian mafia network, where we unveil key complex block structures, mostly hidden from  state-of-the-art alternatives.
\end{abstract}


\end{frontmatter}


\section{Introduction} \label{sec_1}
Network data are ubiquitous in modern applications, and there is recurring interest in block structures defined by groups of nodes that share similar connectivity patterns \citep[e.g.,][]{fortunato2016}.  Our focus is on studying networks of individuals involved in organizing crime.  In this setting, it is of considerable interest to infer shared connectivity patterns among different suspects, based on data provided by investigations, in order to obtain key insights into the hierarchical  structure of criminal organizations \citep[e.g.,][]{campana2016,faust2019,diviak2019,campana2020}.

The relevance of this endeavor has motivated an increasing  shift in modern forensic studies away from classical descriptive analyses of criminal networks 	\citep[e.g.,][]{krebs2002,carley2002,morselli2009,malm2011,agreste2016,grassi2019,cavallaro2020}, and towards studying more complex group structures involving the monitored suspects \citep[e.g.,][]{ferrara2014,calderoni_2014,magalingam2015,calderoni2017,liu2018,sangkaran2020}. These contributions have provided valuable initial insights into the structure and functioning of several criminal organizations.  However, the focus has been on classical community detection algorithms \citep{girvan2002,newman_2004,newman2006,blondel_2008}, which infer groups of criminals characterized by dense within-block connectivity and sparser connections between different blocks \citep{fortunato2016}.  Such approaches are overly simplified and ignore other fundamental block structures, such as core-periphery, disassortative and weak community patterns \citep[e.g.,][]{fortunato2016}.  These more nuanced structures are inherent to criminal organizations, which exhibit an intricate combination of vertical and horizontal hierarchies of block interactions \citep{paoli2007,morselli2007,le2012,catino2014}. Disentangling such complex architectures is fundamental  to  inform preventive and repressive operations. However, this task requires improved methods combined with more realistic representations of criminal networks that incorporate a broader set of recurring block structures, beyond assortative communities.

An initial strategy for addressing the above objectives is to consider spectral clustering algorithms \citep{von2007} and stochastic block models  \citep{Holland_1983,Nowicki_2001}.  Both methods learn more general block architectures in network data and hence, despite their limited use in forensic studies, are expected to unveil criminal structures currently hidden to community detection algorithms. Nonetheless, as clarified in Sections \ref{sec_1.1}--\ref{sec_1.2}, several aspects of criminal network studies still require careful statistical innovations. A crucial one is the coexistence of several community, core-periphery and disassortative architectures whose number, size and structure are unknown and partially obscured by the measurement errors arising from the investigations. To ensure accurate learning  in these challenging settings it is fundamental to rely on an extended, yet interpretable, class of model-based solutions encompassing a variety of flexible mechanisms for the formation of suspect groups. Such processes should also allow structured inclusion of external information and facilitate the adoption of principled methods for estimation, prediction, uncertainty quantification and model selection, within a single realistic modeling framework.


\subsection{The Infinito network}
\label{sec_1.1}
Our motivation is drawn from a large law-enforcement operation, named {\em Operazione Infinito}, that was conducted in Italy from 2007 to 2009 for disentangling and disrupting the core structure of the 'Ndrangheta mafia in Lombardy, north of Italy. According to  the pre-trial detention order produced by the preliminary investigation judge of Milan\footnote{\label{note1}Tribunale di Milano, 2011. Ordinanza di applicazione di misura coercitiva con mandato di cattura --- art. 292 c.p.p. (Operazione Infinito). Ufficio del giudice per le indagini preliminari (in Italian).}, such a criminal organization, also referred to as  {\em La Lombardia}, is a key example of a deeply rooted, highly structured and hard to untangle covert architecture with a disruptive and pervasive impact, both locally and internationally \citep[][]{paoli2007,catino2014}. This motivates our efforts to provide an improved understanding of its hidden hierarchical structures via innovative block-modeling of the relationships among its monitored affiliates.

\begin{figure}[t]
	\centering
\includegraphics[width=12.7cm]{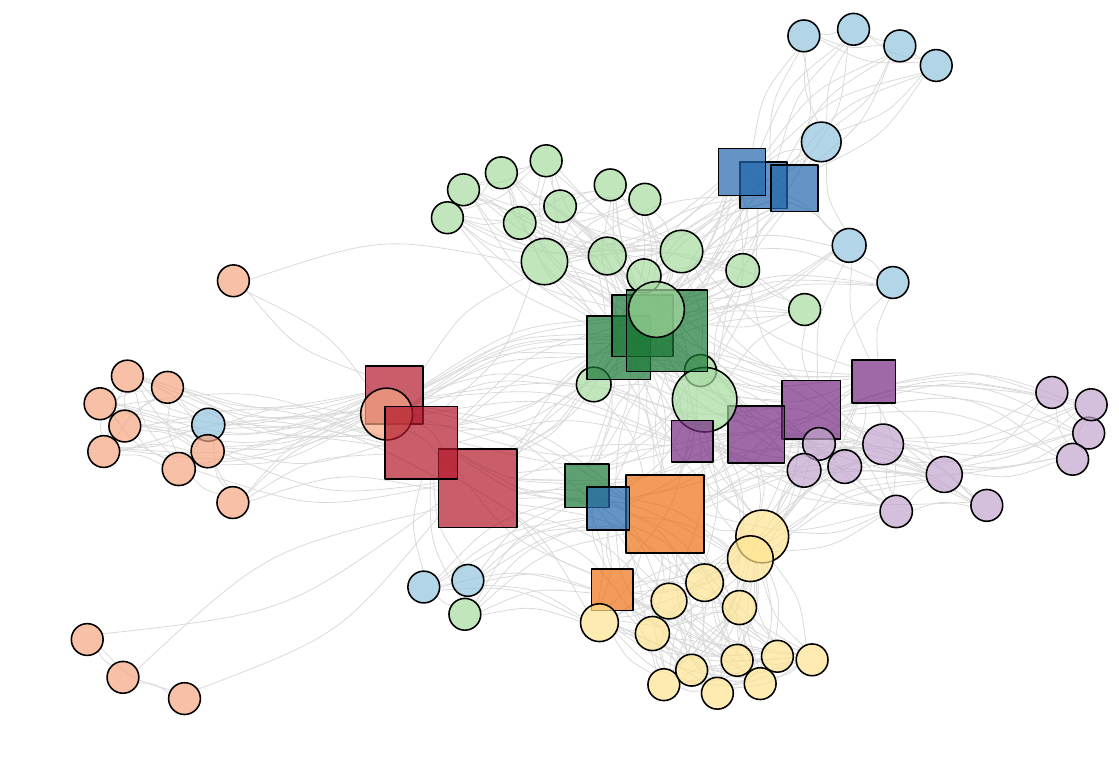}
	\caption{Graphical representation of the Infinito network. Node positions are obtained via force-directed placement \citep{fru1991}. Node size is proportional to the corresponding betweenness, whereas colors indicate the presumed  locale membership. Darker square nodes represent the bosses of each  locale, while lighter circles indicate the affiliates.}
	\label{F1}
\end{figure}

Raw data are available at \url{https://sites.google.com/site/ucinetsoftware/datasets/covert-networks} and comprise information on the co-participation of 156 suspects at 47 monitored summits of the criminal organization, as reported in the judicial acts\textsuperscript{\ref{note1}} that were issued upon request by the prosecution.  Consistent with our main goal of shedding light on the internal structure of  {\em La Lombardia} via inference on the block connectivity patterns among its affiliates, we focus on the reduced set of 118 suspects that attended at least one summit and were classified in the judicial acts as members of this specific criminal organization. Since only  $18\%$ of these affiliate pairs co-attended at least one of the summits and, among them, just $5\%$ co-participated in more than one meeting, we consider here the binary adjacency matrix indicating the presence or absence of  a co-attendance in at least one of the monitored summits. Due to the sparse and almost-binary form of the original counts, this dichotomization leads to a negligible loss of information and is beneficial in reducing the noise that  may arise from investigations of multiple summits. Moreover, because of the  highly regulated  'Ndrangheta coordinating processes \citep{paoli2007,catino2014}, the co-attendance of at least one summit is arguably sufficient to declare the presence of a connection among two affiliates. 

More problematic is the possible presence of false negatives which may arise in such studies as a result of coverting strategies implemented by the criminal organization to carefully balance the tradeoff between efficiency and security \citep{morselli2007}. These covert patterns are not altered by the dichotomization procedure, and further motivate the development of improved methods for principled uncertainty quantification and structured borrowing of information among affiliates via the inclusion of available knowledge on the criminal organization and on suspects' external attributes. For example, current forensic theories \citep[e.g.,][]{paoli2007,catino2014} and initial quantitative analyses \citep[e.g.,][]{calderoni_2014,calderoni2017} suggest that the internal organization of 'Ndrangheta revolves around specific blood family relations, which may be further aggregated at the territorial level in structural coordinated units, named {\em locali}. Each  {\em locale} controls a specific territory, and has a further layer of regulated hierarchy defined by a group of affiliates, and comparatively fewer bosses that are in charge of leading the {\em locale}, managing the funds, overseeing violent actions and guaranteeing the communication flows. Information on presumed {\em locale} membership and role can be retrieved, for each suspect of interest, from the judicial acts\textsuperscript{\ref{note1}} of {\em Operazione Infinito} and, as shown in the graphical representation of the {\em Infinito network} in Figure~\ref{F1}, could help in assisting inference on the hidden underlying block structure, thus reducing the impact of coverting strategies.

The inclusion of the aforementioned node attributes motivates careful and principled probabilistic representations accounting for the fact that these sources of  external information are produced by an investigation process and, therefore,  may be prone to measurement errors. Despite its relevance and potential benefits, this endeavor has been largely neglected in the analysis of criminal networks and, as discussed in Section \ref{sec_1.2}, state-of-the-art methods for block-modeling lack a general solution to include error-prone node attribute effects in the partition process and quantify the magnitude of the improvements relative to no supervision. To cover this gap and flexibly learn the complex variety of block structures in noisy criminal networks, we develop a  novel class of extended stochastic block models (\textsc{esbm}) that formally quantify uncertainty in the suspects' grouping structure --- including in the number, size and composition of the blocks --- via  Gibbs-type priors \citep{Gnedin2005,Lijoi2007a,lijoi2007b, lijoi2008} for the underlying partition; see also \citet{de_2013} for a recent review. 

As clarified in Section \ref{sec_2}, although the block-modeling literature has focused on a much less extensive set of processes, the Gibbs-type class is well motivated in providing broad,   interpretable and realistic probabilistic generative mechanisms for the formation of suspects' groups. This allows careful incorporation of probabilistic structure within a single modeling framework which is amenable to novel extensions for the inclusion of probabilistic homophily with respect to error-prone external attributes, and for careful model-based inference  on the  partition structure via refined methods for estimation, uncertainty quantification, model selection and prediction. To assess out-of-sample predictive performance, we perform inference on the $V=84$ suspects affiliated to the $5$  most populated {\em locali}, and hold out as a test set the $34$ members of those smaller-sized {\em locali} with $\leq 6$ monitored affiliates. As discussed in \citet{calderoni_2014}, such a choice is also beneficial in reducing potential issues arising from the  incomplete identification of low-sized {\em locali} during investigations, and, due to the modular organization of   'Ndrangheta  \citep[e.g.,][]{paoli2007,catino2014}, it arguably leads to a more accurate learning of its core recurring hierarchies.


\subsection{Relevant literature}
\label{sec_1.2}
 The relevance of  learning block structures in networks has motivated a collective effort by various disciplines towards the development of methods for detecting node groups, ranging from algorithmic strategies \citep{girvan2002,newman_2004,newman2006,von2007,blondel_2008} to model-based solutions \citep{Holland_1983,Nowicki_2001,kemp_2006,air_2008,karrer2011,athreya2017,geng_2019}; see  \citet{fortunato2016}, \citet{abbe2017}, and  \citet{lee2019} for an overview. 
 
\begin{figure}[t!]
	\centering
\includegraphics[width=16.1cm]{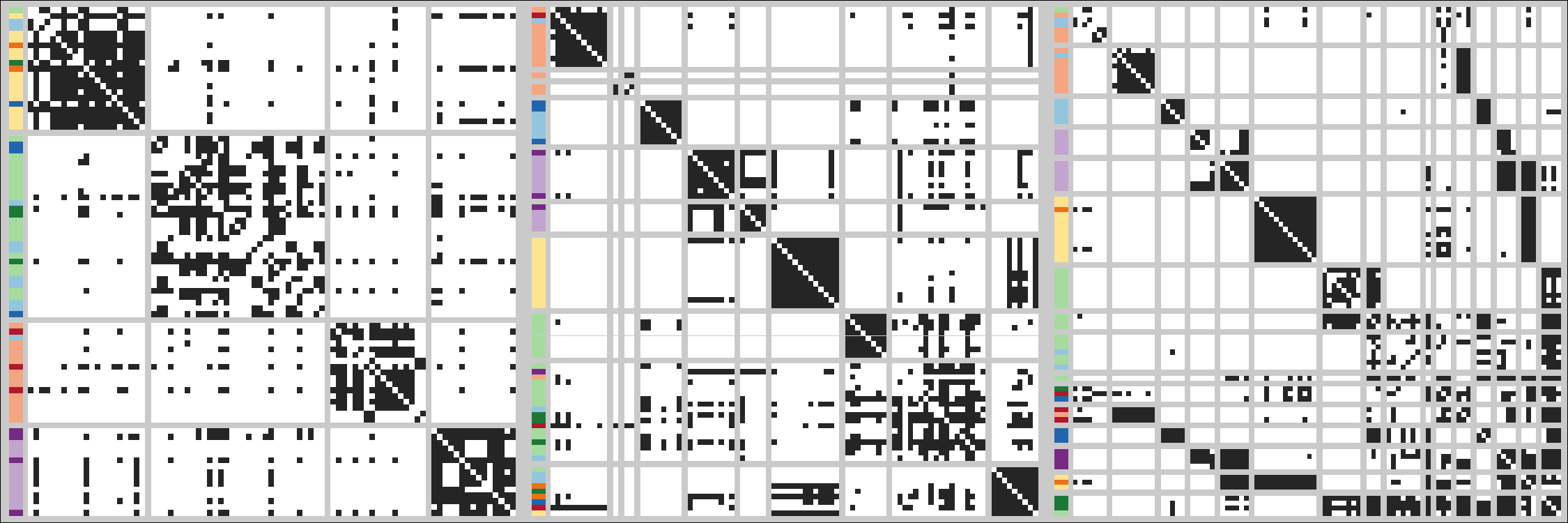}
		\put(-82,-15){(c)}
		\put(-235,-15){(b)}
		\put(-390,-15){(a)}
	\caption{Adjacency matrix of the  Infinito network with nodes re-ordered and partitioned in blocks according to the clustering structure estimated under three different methods: (a) community detection via the Louvain algorithm \citep{blondel_2008}; (b) spectral clustering \citep{von2007} with the number of groups obtained via a combination of the model selection procedures in the \texttt{R} package \texttt{randnet}; (c) \textsc{esbm} with supervised Gnedin process prior.  Black and white cells represent edges and non-edges, respectively. Side colors correspond to the different  locali, with darker and lighter shades denoting  bosses and affiliates, respectively. }
	\label{F2}
\end{figure}

 Despite being  routinely implemented in criminal network studies, most algorithmic approaches focus on detecting communities characterized by a dense connectivity within each block and  sparser connections among different blocks \citep{girvan2002,newman_2004,newman2006,blondel_2008}. This constrained search is expected to provide a limited and possibly biased view of the key modules that are hidden in criminal networks. For instance, Figure~\ref{F1} clearly highlights a core-periphery structure underlying the {\em Infinito network}, with communities of affiliates in peripheral positions and groups of bosses at the core.  According to panel (a) in Figure~\ref{F2},  state-of-the-art algorithms for community detection \citep{blondel_2008} applied to the {\em Infinito network} obscure such patterns by  over-collapsing some {\em locali}, while failing to separate affiliates from bosses.

These issues motivate  focus on alternative solutions aimed at grouping nodes which are characterized by common connectivity patterns within the network, rather than just exhibiting community structures. One possibility to address this goal from an algorithmic  perspective is to rely on spectral clustering \citep{von2007}. This strategy accounts for general block structures  and possesses desirable properties, including consistency in estimation of the partition structure underlying various model-based representations \citep{rohe2011,sussman2012,sarkar2015,lei2015,athreya2017,zhou_2019}. As shown in panel (b) of Figure~\ref{F2}, this yields improvements in learning complex block structures within the {\em Infinito network}  relative to classical community detection algorithms. Nonetheless, spectral clustering lacks extensive methods for inference beyond point estimation, requires pre-specification or heuristic algorithms to choose the unknown number of groups, and faces practical instabilities. As a result, this strategy is suboptimal relative to carefully chosen model-based approaches; see panel (c) in Figure~\ref{F2} for an example of the gains that can be obtained over spectral clustering by the methods developed in Sections~\ref{sec_2}--\ref{sec_3}.

Among the generative models for learning groups of nodes in network data, the stochastic block model (\textsc{sbm}) \citep{Holland_1983,Nowicki_2001} is arguably the most widely implemented and well-established formulation, owing also to its  balance among simplicity and flexibility \citep{abbe2017,lee2019}. In \textsc{sbm}s, the probability of an edge only depends on the cluster memberships of the two involved nodes, thus allowing efficient inference on node groups and on block probabilities --- which can also characterize  disassortative, core-periphery or weak community patterns,  and combinations of such structures \citep{fortunato2016}. These desirable properties have motivated extensive theory \citep{zhao2012,bickel2013,olhede2014,noroozi2020} and various generalizations of the \textsc{sbm} \citep{tallberg_2004,kemp_2006,handcock2007,air_2008,karrer2011, schmidt_2013,newman2016,white2016,sengupta2018,rastelli2018,geng_2019,stanley2019,fosdick2019}.

Part of these extensions aim at addressing two relevant open issues with classical  \textsc{sbm}s, that also arise in criminal networks. First, in real-world applications the number of underlying groups is typically not known and has to be inferred from the data. Therefore, classical \textsc{sbm} formulations  based on a fixed and pre-specified number of groups \citep{Holland_1983,Nowicki_2001} are conceptually unappealing in precluding uncertainty quantification on the unknown number of non-empty clusters, while state-of-the-art model selection procedures for choosing this quantity \citep[][]{le2015,saldana2017,wang2017,chen2018,li2020network} led to mixed and biased results when applied to realistic criminal networks in the simulation studies reported in Section~\ref{sec_4}.  The second important problem is that, as discussed in Section \ref{sec_1},  it is common to observe external and possibly error-prone node attributes that may effectively inform the grouping mechanism.  Hence, \textsc{sbm}s require extensions to include such information in the partitioning process. 

A successful answer to the first open issue has been provided by Bayesian nonparametric solutions replacing the original Dirichlet-multinomial process for node partitioning  \citep{Nowicki_2001} with alternative priors that allow the number of groups to grow adaptively with the network size via  
the Chinese restaurant process (\textsc{crp}) \citep{kemp_2006, schmidt_2013} or to be finite and random under a mixture-of-finite-mixtures representation \citep{geng_2019}. Unfortunately, all these extensions have been developed separately and \textsc{sbm}s still lack a unifying framework, which would be conceptually and practically useful to clarify common properties, develop broad computational and inferential strategies,  and identify novel solutions that may effectively address the problems arising from the criminal network  discussed in Section~\ref{sec_1.1}. To address this gap, we unify in Section~\ref{sec_2} most of the aforementioned formulations within an extended stochastic block model (\textsc{esbm}) framework based on Gibbs-type priors. 

As clarified in Section~\ref{sec_2.2.2}, this broad family of prior distributions also  allows the natural inclusion of error-prone node attributes in a principled manner via product partition models (\textsc{ppm}s) \citep{Hartigan1990,Quintana2003, Lijoi2007a, muller2011product} which favor the formation of groups that are homogenous with respect to attributes, thereby incorporating probabilistic homophily.  This property is known to play a key role in the formation of blocks within networks, and has inspired modifications of algorithmic strategies to learn group patterns coherent not only with network structure but also with pairwise similarities among node attributes \citep[e.g.,][]{zhang2016,binkiewicz2017}. These solutions often yield to practical gains, but inherit the inferential limitations of the unsupervised counterparts. Available model-based strategies  \citep[e.g.,][]{tallberg_2004,gormley2010,kim_2012,newman2016,white2016,zhao2017} commonly treat node membership variables as categorical responses whose formation depends on attributes via a higher-level regression model which, however, does not explicitly incorporate measurement errors in the attributes. Closer to the methods developed in Section~\ref{sec_2.2.2} are mixture representations defining a joint model for  the network and the node attributes, under the assumption of a shared underlying partition  that influences the formation of both data structures \citep[e.g.,][]{xu2012,yang2013,stanley2019}. These models arise as attribute-assisted versions of the original \textsc{sbm} by  \citet{Nowicki_2001}, but lack a broader modeling, inferential and computational framework  to incorporate and compare more general priors on the random partition, beyond the Dirichlet-multinomial. Section~\ref{sec_2.2.2} covers this gap by leveraging the connection between  Gibbs-type priors and \textsc{ppm}s, which further provides a direct and principled characterization of homophily. 

Within the Gibbs-type class, we will mainly focus on the Gnedin process \citep{Gnedin2010, de_2013b} as an example of prior which has not yet been employed in \textsc{sbm}s, but exhibits analytical tractability, desirable properties, theoretical guarantees and promising empirical performance in applications; see panel (c) in Figure~\ref{F2}. As clarified in Section~\ref{sec_3}, our framework allows posterior computation via an easy-to-implement collapsed Gibbs sampler, and motivates general strategies for uncertainty quantification, prediction and model assessment, thus fully exploiting the advantages of a model-based approach over algorithmic strategies. The performance of key priors within the \textsc{esbm} class and the magnitude of the improvements relative to state-of-the-art competitors are illustrated in  Section~\ref{sec_4} with extensive simulations focusing on realistic criminal network structures. In light of these results, we opt for a supervised Gnedin process to analyze the {\em Infinito network} in Section~\ref{sec_5}, obtaining a novel in-depth view of the modular organization of 'Ndrangheta that was hidden to previous quantitative studies. Concluding remarks are provided in Section~\ref{sec_6}, where we also mention possible extensions to degree-corrected  stochastic block models \citep{karrer2011} and mixed membership stochastic block models \citep{air_2008}. Codes and data to reproduce all our results are available at \url{https://github.com/danieledurante/ESBM}.


\section{Extended stochastic block models}
\label{sec_2}
Consider a binary undirected network with $ V $ nodes, and let $ \bY $ denote its $ V\times V $ symmetric adjacency matrix, with elements $ y_{vu}=y_{uv}=1 $ if nodes $ v $ and $ u $ are connected, and $ y_{vu}=y_{uv}=0 $ otherwise. In our criminal network application self-loops are not allowed and, hence, are not included in the generative model. In Section~\ref{sec_2.1}, we first present the statistical model relying on classical \textsc{sbm} representations, and then characterize, in Section~\ref{sec_2.2.1}, the prior on the node partition via Gibbs-type processes leading to our general  \textsc{esbm} class. Such a unified representation is further extended in Section~\ref{sec_2.2.2} to include information from error-prone node attributes. Consistent with our motivating application, we focus on binary undirected edges and categorical attributes, but our approach can be naturally extended to other types of networks and covariates, as highlighted in the final discussion.


\subsection{Model formulation}
\label{sec_2.1}
 \textsc{sbm}s \citep{Holland_1983,Nowicki_2001} partition the nodes into $H$ mutually exclusive and exhaustive groups, with nodes in the same cluster sharing  common connectivity patterns. More specifically, \textsc{sbm}s assume that the sub-diagonal entries $ y_{vu} $, for $v=2, \ldots, V$, $u=1, \ldots, v-1$, of the symmetric adjacency matrix $ \bY $ are conditionally independent Bernoulli random variables with associated probabilities $\theta_{z_v,z_u} \in (0,1)$ depending only on the group memberships $z_v$ and $z_u$ of the two involved nodes $ v $ and $ u $. Let $\bz=(z_1, \ldots, z_V)^{\intercal} \in \{1, \ldots, H\}^V$ be the node membership vector associated to the generic node partition $\{Z_1,\ldots,Z_{H}\} $, so that  $z_v = h$  if and only if $v \in Z_h$, and denote with $ \bTheta $ the $ H \times H $ symmetric matrix whose generic element $ \theta_{hk} \in (0,1)$ denotes the probability of an edge between a node in group $ h $ and a node in group $ k $. Then, the likelihood for $ \bY $ is $p(\bY \mid  \bz, \bTheta)= \prod\nolimits_{h=1}^{H}\prod\nolimits_{k=1}^h \theta_{hk}^{m_{hk}}(1-\theta_{hk})^{\overline{m}_{hk}},$ where $m_{hk}$ and $\overline{m}_{hk}$ denote the number of edges and non-edges between nodes in groups $h$ and $k$, respectively. 

Classical  \textsc{sbm}s \citep{Holland_1983,Nowicki_2001} assume independent $\mbox{Beta}(a,b) $ priors for the block probabilities $\theta_{hk}$. Therefore,  the joint density for the diagonal and sub-diagonal elements of  $\bTheta$ is $p(\bTheta) = \prod\nolimits_{h=1}^{H} \prod\nolimits_{k=1}^h  [\theta_{hk}^{a-1}(1-\theta_{hk})^{b-1}]\mbox{B}(a,b)^{-1},$ where $\mbox{B}(\cdot,\cdot)$ is the Beta function. Although quantifying prior uncertainty in the block probabilities is important, the overarching goal in \textsc{sbm}s is to infer the node partition. Consistent with this focus,  $ \bTheta $~is commonly treated as a nuisance parameter which is marginalized out in $p(\bY \mid \bz, \bTheta)$ via beta-binomial conjugacy, obtaining
\begin{eqnarray}
\label{eq_marg_lhd}
p(\bY \mid \bz) = \prod\nolimits_{h=1}^{H} \prod\nolimits_{k=1}^h \frac{\mbox{B}(a+m_{hk}, b+ \overline{m}_{hk})}{\mbox{B}(a,b)}.
\end{eqnarray}
As we will clarify in Section~\ref{sec_3}, this marginalization is also useful for computation and inference. The likelihood in~\eqref{eq_marg_lhd} is common to several \textsc{sbm} extensions, which then differ in the choice of the probabilistic mechanism underlying  $\bz$. Let $\overline{H} \geq H$ be the total number of possible groups in the whole population of nodes, and denote with $\bar{\bz}=(\bar{z}_1, \ldots, \bar{z}_V)^{\intercal} \in \{1, \ldots, \overline{H}\}^V$ the indicators of the population clusters for the $V$ observed nodes. A natural option to define the generative process for the partition is to consider a Dirichlet-multinomial prior distribution for $\bar{\bz}$, obtained by marginalizing the vector of group probabilities $ \bpi = (\pi_1,\ldots,\pi_{\overline{H}})\sim \mbox{Dirichlet}(\bbeta)$ out of a multinomial likelihood for  $\bar{\bz}$, in which  $\mbox{pr}(\bar{z}_v=h \mid \bpi) = \pi_h$ for $v=1, \ldots, V$. If $ \overline{H}$ is fixed and finite, this leads to the original  Bayesian \textsc{sbm} \citep{Nowicki_2001}.  However, as already discussed, the number of groups in criminal networks is usually unknown and has to be inferred from the  data. A possible solution consists in placing a prior on $\overline{H} $, which leads to the mixture-of-finite-mixtures (\textsc{mfm}) version of the \textsc{sbm} in \citet{geng_2019}.  Another option is a Dirichlet process partition mechanism, corresponding to the infinite relational model~\citep{kemp_2006}. Such an infinite mixture model differs from \textsc{mfm} in that $ \overline{H}=\infty$, meaning that infinitely many nodes would give rise to infinitely many groups. Note that the total number of possible clusters $ \overline{H} $ should not be confused with the number of occupied clusters $ H $. The latter is defined as the number of distinct labels in $ \bar{\bz} $, and is upper bounded by $ \min\{V,\overline{H}\} $. 

Notably, all  the above solutions are specific examples of Gibbs-type priors \citep[e.g.,][]{de_2013}, thus motivating our unified \textsc{esbm} class presented in Section \ref{sec_2.2}. Before introducing this extension it is worth noticing that $\bar{\bz}$ identifies labeled clusters. Hence, a vector $\bar{\bz}$ and its relabelings are regarded as distinct objects, even though they identify the same partition. Throughout the rest of the paper we will rely on the previously-defined vector $ \bz $, which denotes all relabelings of $\bar{\bz}$ that lead to the same partition. For simplicity, we assume that $z_v \in  \{1,\dots,H\}$, which corresponds to avoiding empty groups. This does not modify likelihood~\eqref{eq_marg_lhd}, which is invariant under relabeling; i.e.,  $ p(\bY \mid \bz)=p(\bY \mid \bar{\bz}) $. 


\subsection{Prior specification}
\label{sec_2.2}
As illustrated in Section~\ref{sec_2.1}, several priors for $\bz$  have been considered in the context of \textsc{sbm}s, including the Dirichlet-multinomial \citep{Nowicki_2001}, the Dirichlet process \citep{kemp_2006}, and mixtures of finite Dirichlet mixtures \citep{geng_2019}. Interestingly, these are all examples of Gibbs-type priors, which stand out for their analytical and computational tractability; see \citet{de_2013} for a comprehensive review. In Section~\ref{sec_2.2.1} we propose the \textsc{esbm} as a unifying framework characterized by the choice of a Gibbs-type prior for $\bz$. This formulation includes the previously-mentioned \textsc{sbm}s as special cases and offers new alternatives by exploring the whole Gibbs-type class and its relation with \textsc{ppm}s \citep{Hartigan1990,Quintana2003,Lijoi2007a}. This connection with \textsc{ppm}s is exploited in Section~\ref{sec_2.2.2} to supervise the prior via possibly error-prone node attributes.


\subsubsection{Unsupervised Gibbs-type priors}
\label{sec_2.2.1}
Gibbs-type priors  are defined on the space of unlabeled  group indicators $\bz$. For $ a>0 $, denote the ascending factorial with $(a)_n = a(a+1)\cdots(a + n - 1)$ for any $ n \geq 1 $, and set $(a)_0 = 1$. A probability mass function $ p(\bz) $ is of Gibbs-type if and only if 
\begin{eqnarray} \label{eq_eppf}
p(\bz) = \mathcal{W}_{V,H} \prod\nolimits_{h=1}^{H} (1 - \sigma)_{n_h-1},
\end{eqnarray}
where $n_h$ is the number of nodes in cluster $ h $, $ \sigma <1 $ denotes the so-called \textit{discount parameter} and $ \{ \mathcal{W}_{V,H}: 1\leq H \leq V \} $ is a collection of non-negative weights satisfying the recursion $\mathcal{W}_{V,H} = (V - H \sigma) \mathcal{W}_{V+1,H} + \mathcal{W}_{V+1,H+1}$, with $ \mathcal{W}_{1,1} = 1$. As shown in \citet{Lijoi2007a}, the class of random partitions induced by Gibbs-type priors coincides with exchangeable  \textsc{ppm}s,  which are probability models for random partitions $\bz$ of the form $p(\bz) \propto  c(Z_1) \cdots c(Z_{H})$, where $\{Z_1,\ldots,Z_{H}\} $ is the partition associated to $\bz$, whereas $ c(\cdot) $ is a non-negative  \emph{cohesion function} measuring the homogeneity within each cluster. Such a connection will be useful to incorporate node-specific attributes in \textsc{esbm}s. Interestingly, Gibbs-type priors represent a broad, yet tractable, class whose predictive distribution \citep{lijoi2007b} implies that membership indicators $\bz$ can be obtained in a sequential and interpretable manner according to
\begin{eqnarray} \label{eq_urn1}
\mbox{pr}(z_{V+1} = h \mid \bz) \propto \begin{cases} \mathcal{W}_{V+1,H} (n_h-\sigma) & \text{for} \ h=1, \ldots, H, \\ \mathcal{W}_{V+1,H+1} & \text{for} \ h= H + 1. \\ \end{cases} 
\end{eqnarray}
Hence, the group assignment process can be interpreted as a simple seating mechanism in which a new node is assigned to an existing cluster $h$ with probability proportional to the current size $n_h$ of that cluster, discounted by a global factor $\sigma$ and further rescaled by a weight  $\mathcal{W}_{V+1,H}$, which may depend both on the size $V$ of the network and on the current number $H$ of non-empty groups. Alternatively, the incoming node is assigned to a new cluster with probability proportional to $\mathcal{W}_{V+1,H+1}$. Such a general mechanism is conceptually  appealing in our application to criminal networks since it realistically accounts for group sizes $n_h$, network size $V$ and complexity $H$ in the formation process of the modular structure underlying the criminal organization, while providing a variety of possible generative mechanisms under a single modeling framework. In the examples below we show how commonly used priors in \textsc{sbm}s and unexplored alternatives of interest in criminal network studies can be obtained as special cases of~\eqref{eq_urn1}.

\begin{example}[\textsc{dm} -- Dirichlet-multinomial]
\label{ex_dm} 
Let $ \sigma < 0 $ and consider the collection of weights  $ \mathcal{W}_{V,H} = [\beta^{H-1}/(\beta \overline{H} {+}1)_{V-1}] \prod_{h=1}^{H-1}(\overline{H}-h) \mathbbm{1}(H \le \overline{H}) $ for some $ \beta = -\sigma $ and $\overline{H} \in \{1,2,\dots\}$. Then \eqref{eq_urn1}~coincides with the \textsc{dm} urn-scheme: $\mbox{pr}(z_{V+1} = h \mid \bz) \propto n_h + \beta$  for $h=1, \ldots, H$ and $\mbox{pr}(z_{V+1} = H + 1 \mid \bz) \propto \beta (\overline{H} - H) \mathbbm{1}(H \le \overline{H})$.
\end{example}

\begin{example}[\textsc{dp} -- Dirichlet process] 
\label{ex_dp}
Let $\sigma = 0$ and set $\mathcal{W}_{V,H} = \alpha^{H}/(\alpha)_V$ for some $\alpha > 0$. Then~\eqref{eq_urn1} leads to a \textsc{crp} urn-scheme: $\mbox{pr}(z_{V+1} = h \mid \bz) \propto n_h$  for $h=1, \ldots, H$ and $\mbox{pr}(z_{V+1} = H + 1 \mid \bz) \propto \alpha$. The \textsc{crp} can also be obtained as a limiting   \textsc{dm} with $\beta = \alpha / \overline{H}$, as ${ \overline{H} \rightarrow \infty}$.
\end{example}

\begin{example}[\textsc{py} -- Pitman-Yor process] 
\label{ex_py}
Let $\sigma \in [0,1)$ and set $\mathcal{W}_{V,H} = [\prod_{h=1}^{H-1}(\alpha {+} h\sigma)]/(\alpha + 1)_{V-1}$ for some $\alpha > - \sigma$. Then \eqref{eq_urn1}~characterizes the \textsc{py} process: $\mbox{pr}(z_{V+1} = h \mid \bz) \propto n_h - \sigma$  for $h=1, \ldots, H$ and $\mbox{pr}(z_{V+1} = H + 1 \mid \bz) \propto \alpha + H\sigma$. This scheme reduces to a \textsc{dp} when $\sigma = 0$.
\end{example}

\begin{example}[\textsc{gn} -- Gnedin process]
\label{ex_gnedin} 
Let $\sigma = -1$ and define $\mathcal{W}_{V,H} = [(\gamma)_{V-H} \prod_{h=1}^{H-1}(h^2 - \gamma h)] / \prod_{v=1}^{V-1}(v^2 + \gamma v)$ for some $\gamma \in (0,1)$. Then~\eqref{eq_urn1} identifies the \textsc{gn} process: $\mbox{pr}(z_{V+1} = h \mid \bz) \propto (n_h + 1)(V - H + \gamma)$  for $h=1, \ldots, H$ and $\mbox{pr}(z_{V+1} = H + 1 \mid \bz) \propto H^2 - H\gamma$.
\end{example}

 Other  popular examples of tractable Gibbs-type priors can be found in \cite{Lijoi2007a,lijoi2007b,de_2013b,de_2013} and \citet{miller_2018}.
 
 Priors~\textsc{dm}, \textsc{dp}, \textsc{py} and \textsc{gn} provide various realistic generative mechanisms for the grouping structure in criminal networks, thus allowing analysts to choose the most suitable one  for a given study, or possibly test different  specifications under a single modeling framework. For example, \textsc{dp} and \textsc{py}  \citep{kemp_2006} may provide useful constructions in the analysis of relatively unstable and  fragmented criminal organizations, such as terrorist networks, which are characterized by multiple small cells and even {\em lone wolves}. As shown in Table~\ref{tab_1}, when the growth is expected to be rapid, i.e., $\mathcal{O}(V^\sigma)$, and  possibly favoring the formation of low-sized groups, \textsc{py} may be a more sensible choice relative to \textsc{dp}, which  in turn would be recommended in regimes with slower increments, i.e., $ \mathcal{O}(\log V) $. Organized crime, such as 'Ndrangheta, is instead characterized by a more stable and highly regulated  modular architecture which might support the use of priors with a finite number $\overline{H}$ of population clusters, such as \textsc{dm} \citep{Nowicki_2001} and \textsc{gn}. Clearly, in most forensic studies, $\overline{H}$ is  unknown and, hence, quantifying uncertainty in $\overline{H}$ under \textsc{gn} provides a more realistic choice than fixing $\overline{H}$ as in  \textsc{dm}.  In fact, the  \textsc{gn} process can be derived from \textsc{dm} by placing a prior on $\overline{H}$, thus making it random. Specifically, the distribution  $p_\textsc{gn}(\bz)$ of $\bz$ under the \textsc{gn}  process  can be  expressed as
\begin{eqnarray*}
p_\textsc{gn}(\bz) = \sum\nolimits_{h=1}^\infty \mbox{pr}_\textsc{gn}(\overline{H} = h) p_\textsc{dm}(\bz; 1, h),
\end{eqnarray*}
where $p_\textsc{dm}(\bz{;} 1, h)$ is the Dirichlet-multinomial distribution  in the first Example, with $\beta=1$ and $\overline{H}=h$, whereas $\mbox{pr}_\textsc{gn}(\overline{H} = h) = \gamma(1-\gamma)_{h-1}/h!$ can be interpreted as the prior  on $\overline{H}$ under  \textsc{gn}.  Although different prior choices for $\overline{H}$ might be considered \citep{de_2013,miller_2018,geng_2019}, the  \textsc{gn} process has conceptual and practical advantages in applications to criminal networks. First, the sequential mechanism described in the fourth Example has a simple analytical expression that facilitates posterior inference and prediction. Moreover, the distribution $\mbox{pr}_\textsc{gn}(\overline{H} = h) = \gamma(1-\gamma)_{h-1}/h!$ has the mode at $1$, heavy tail and infinite expectation \citep{Gnedin2010}. Hence, the associated \textsc{mfm} favors parsimonious representations of the block structure in criminal organizations which facilitate repressive operations, but preserves robustness to  $\overline{H}$ due to heavy-tails.

\begin{table}[t]
	\centering
	\caption{A classification of Gibbs-type priors.}
	\begin{tabular}{lllll}
	\toprule
		 & ${\bar H}$ & $\sigma$  &  $H$ (growth) & Example  \\
		\midrule
		\textsc{I} & Fixed & $\sigma < 0$ & -- & Dirichlet-multinomial (\textsc{dm}) \\
		\textsc{II} & Random & $\sigma < 0$ & -- & Gnedin process (\textsc{gn}) \\
		\textsc{III}.a & Infinite & $\sigma = 0$  & $\mathcal{O}(\log{V})$ & Dirichlet process (\textsc{dp})\\
		\textsc{III}.b & Infinite & $\sigma \in (0,1)$  & $\mathcal{O}(V^\sigma)$ & Pitman-Yor process (\textsc{py})\\
	\bottomrule
	\end{tabular}
	 \label{tab_1}
\end{table}

Priors on $\overline{H}$ quantify the uncertainty in the total number of groups that one would expect if $V \rightarrow \infty$. However, in practice, the number of non-empty groups $H$ occupied by the observed $V$ nodes is of more direct interest and can also guide the choice of the prior hyperparameters.   Under Gibbs-type priors this quantity has a closed form probability mass function which coincides with $\mbox{pr}(H = h) =  \mathcal{W}_{V,h}\mathcal{C}(V,h; \sigma)\sigma^{-h}$ for every $h=1,\dots,V$, where $\mathcal{C}(V,h;\sigma)$ denotes the so-called generalized factorial coefficient~\citep{Gnedin2005}. The \textsc{dp} case is recovered when $\sigma \rightarrow 0$. In \url{https://github.com/danieledurante/ESBM} we provide codes to evaluate such quantities under the Gibbs-type priors in Table~\ref{tab_1}, and then exploit these values for guiding the choice of the hyperparameters, a strategy first proposed in \citet{Lijoi2007a,lijoi2007b}. In Sections~\ref{sec_4}--\ref{sec_5} this is accomplished by combining a visual inspection of the prior distribution induced on~$H$ with the analysis of its relevant moments. This strategy provides a practically effective solution in a broad set of applications where expert knowledge can be directly quantified through prior information on $H$, which naturally translates into specific hyperparameters  for the four examples of Gibbs-type priors in Table~\ref{tab_1}.

In addition to its practical relevance, the above result clarifies also the asymptotic behavior of $H$. Indeed, the distribution of $H$ converges to a point mass in scenario \textsc{I}, to a proper distribution in scenario \textsc{II} and to a point mass at infinity in scenario \textsc{III}. For instance, under \textsc{gn} in the fourth Example, we have 
\begin{eqnarray*}
	\mbox{pr}_\textsc{gn}(H = h) =  \binom{V}{h} \frac{(1 - \gamma)_{h-1}(\gamma)_{V-h}}{(1+\gamma)_{V-1}},
	\quad h=1,\dots,V,
	\end{eqnarray*}
and hence the expectation can be easily computed via $\mathbb{E}_\textsc{gn}(H)=\sum\nolimits_{h=1}^V h \cdot \mbox{pr}_\textsc{gn}(H = h)$. Note that $\lim_{V \rightarrow \infty} \mbox{pr}_\textsc{gn}(H = h) = \mbox{pr}_\textsc{gn}(\overline{H} = h) = \gamma(1-\gamma)_{h-1}/h!$.

The prior on $\overline{H}$ induced by \textsc{gn} also ensures  posterior consistency for the estimated grouping structure. This follows from the theory for \textsc{mfm} in \cite{geng_2019}, that actually applies to any \textsc{dm} with prior on $\overline{H}$ supported on all positive integers. In particular, this holds for \textsc{gn}, thus giving further support for the use of such a prior in the motivating criminal network application. Instead, \textsc{dp} and \textsc{py} unsurprisingly lead to inconsistent estimates for $\overline{H}$ if the data are generated from a model with $\overline{H}_0 < \infty$ \citep{miller2014}. Intuitively, this happens because \textsc{dp} and \textsc{py} assume $\overline{H}=\infty$. Hence, we suggest Gibbs-type priors with $\sigma\geq 0$ only if the analyst believes that $\overline{H}_0=\infty$, that is, when the true number of groups is assumed to grow without bound with the number of nodes; see also Sections~\ref{sec_3.2}, \ref{sec_4} and \ref{sec_5} for additional data-driven strategies to select among the different priors via the \textsc{waic} criterion \citep{watanabe2010,watanabe2013}.


\subsubsection{Supervised Gibbs-type priors}
\label{sec_2.2.2}
When node attributes ${\bf x}_v = (x_{v1},\dots,x_{vd})^\intercal$ are available for each $v=1, \ldots, V$, this external information may support inference on block structures, both in term of point estimation and in reduction of posterior uncertainty. As mentioned in Section~\ref{sec_1}, this is particularly relevant in applications to criminal networks where specific block structures could be purposely blurred by coverting strategies and, therefore, inclusion of informative attributes might help in revealing obscured modules. This solution should also account for the fact that node attributes collected in  investigations may be error-prone.

One option to address  the above goals in a principled manner within \textsc{esbm}s  is to rely on the \textsc{ppm} structure of Gibbs-type priors. Adapting results in \cite{Park2010} and in \cite{muller2011product} to our network setting, this solution is based on the idea of replacing~\eqref{eq_eppf} with
\begin{eqnarray} \label{eq_ppmX}
p(\bz \mid \bX)  \propto \mathcal{W}_{V,H} \prod\nolimits_{h=1}^{H} p(\bX_h)(1 - \sigma)_{n_h-1},
\end{eqnarray}
where $ \bX = ({\bf x}_1,\dots, {\bf x}_V)^\intercal$, whereas  $ \bX_h = \{{\bf x}_v : z_v = h\}$ are the attributes for the nodes in cluster $ h $. In \eqref{eq_ppmX}, $p(\bX_h)$ controls the contribution of $\bX$ to the cluster cohesion by favoring groups that are homogeneous with respect to attribute values, while including uncertainty in the observed attributes. Motivated by the application to the {\em Infinito network}, we consider the case in which each node attribute ${\bf x}_v = x_v \in \{ 1,\ldots,C \} $ is a single categorical variable denoting a suitable combination between  {\em locale} affiliation and role in the criminal organization. This is a common setting in  criminal network studies, where node attributes often come in the form of exogenous partitions defined by the forensic agencies as a result of the investigation process. In these categorical settings, the recommended practice within the \textsc{ppm} framework \citep{muller2011product} is to rely on the Dirichlet-multinomial (without multinomial coefficient) cohesion 
\begin{eqnarray} \label{eq_g}
p(\bX_h) \propto 
\frac{1}{\Gamma(n_h+\alpha_0)} \prod\nolimits_{c=1}^{C} \Gamma(n_{hc}+\alpha_c),
\end{eqnarray}
where $n_{hc}$ is the number of nodes in cluster~$ h $ with attribute value~$ c $, and $ \alpha_0 = \sum_{c=1}^{C} \alpha_c $, with $\alpha_c>0$ for $c=1, \ldots, C$. Including this cohesion function in equation \eqref{eq_ppmX} leads to the following urn scheme 
\begin{eqnarray}
 \mbox{pr}(z_{V+1} = h \mid  \bX,x_{V+1},\bz) \propto \begin{cases}  \frac{n_{hx_{V+1}}+\alpha_{x_{V+1}}}{n_h+\alpha_0} \mathcal{W}_{V+1,H} (n_h-\sigma) & \text{for} \ h=1, \ldots, H, \\  \frac{\alpha_{x_{V+1}}}{\alpha_0} \mathcal{W}_{V+1,H+1} & \text{for} \ h= H + 1, \\ \end{cases} 
 \label{eq_urn2}
\end{eqnarray}
where $ n_{hx_{V+1}} $ is the number of nodes in cluster $ h $ with the same covariate value $c=x_{V+1}$ as node $V+1$, $ n_h$ is the total number of nodes in cluster $ h $, whereas $\alpha_{x_{V+1}}$ is the parameter associated with the category $c=x_{V+1}$ of node $V+1$. As shown in  \eqref{eq_urn2}, the introduction of a $p(\bX_h)$, defined as in \eqref{eq_g}, induces a probabilistic homophily structure which  favors the attribution of a new node  to those groups containing a higher fraction of existing nodes with its same attribute value. 

Besides including realistic homophily structures, the above representation effectively accounts for  possible noise in the attributes. Indeed, the expression for $p(\bX_h)$  in \eqref{eq_g} coincides with the marginal likelihood for the attributes of the nodes in group $h$ under the assumption that the model underlying these quantities is defined by a multinomial with group-specific class probabilities $\bnu_h=(\nu_{1h}, \ldots, \nu_{Ch})^{\intercal}$, which are assigned a  Dirichlet prior with parameters $\balpha=(\alpha_1, \ldots, \alpha_C)^{\intercal}$. Under this interpretation, the supervised Gibbs-type prior in equation~\eqref{eq_ppmX} can be  re-expressed as $p(\bz \mid \bX)  \propto [\mathcal{W}_{V,H} \prod\nolimits_{h=1}^{H} (1 - \sigma)_{n_h-1}]\prod\nolimits_{h=1}^{H} p(\bX_h)\propto p(\bz)p( \bX \mid \bz)$, where $p(\bz)$ is the unsupervised Gibbs-type prior in Section~\ref{sec_2.2.1}, whereas $p( \bX \mid \bz)$ is the likelihood induced by the Dirichlet-multinomial model for the observed node attributes. Hence, learning block structures in $\bY$ under the supervised Gibbs-type prior can be interpreted as a two-step Bayesian procedure in which the unsupervided prior on $\bz$ is first updated with the likelihood for the attributes in $\bX$, and then such a first-step posterior enters as a new prior in the second step to be updated with the information from the observed network  $\bY$. Under the assumption of conditional independence between $\bY$ and $\bX$ given $\bz$, such a two-step process yields the actual posterior for $\bz$, since $p(\bz \mid \bY,\bX) \propto [p(\bz)p(\bX \mid \bz)]p(\bY \mid \bz) \propto p(\bz \mid \bX)p(\bY \mid \bz)$. 

As mentioned in Section~\ref{sec_1.2}, while the induced joint model for $\bY$ and $\bX$ is reminiscent of earlier constructions \citep{xu2012,yang2013,stanley2019}, our solution crucially extends these ideas to  the whole \textsc{esbm} class, well beyond the original Bayesian \textsc{sbm} by  \citet{Nowicki_2001}.


\section{Posterior computation and inference}\label{sec_3}

In Section~\ref{sec_3.1}, we derive a collapsed Gibbs sampler that holds for the whole \textsc{esbm} class presented in Section~\ref{sec_2}. Then, in Section~\ref{sec_3.2} we provide  extensive tools not only for point estimation of the group structure, but also for uncertainty quantification, model selection and prediction. Despite their relevance in routine studies including, for example, the  {\em Infinito network}  motivating application in Section~\ref{sec_1.1}, these aspects have been  partially neglected in the \textsc{sbm} literature.  


\subsection{Collapsed Gibbs sampler}\label{sec_3.1}

The availability of the urn schemes \eqref{eq_urn1} and~\eqref{eq_urn2} for the whole class of Gibbs-type priors allows the derivation of a general collapsed Gibbs sampler that holds for any \textsc{esbm}; see Algorithm~\ref{alg_esbm}. At every iteration, this routine samples the group assignment of each node $ v=1, \ldots, V $ from its full conditional distribution given the adjacency matrix $ \bY $ and the vector $\bz_{-v} $ of the cluster assignments of all the other nodes, excluding $v$. By direct application of the Bayes rule, these full conditional probabilities are 
\begin{eqnarray} \label{eq_exact_full}
\mbox{pr}(z_v=h \mid \bY, \bX, \bz_{-v}) \propto \mbox{pr}(z_v=h \mid \bX, \bz_{-v})  \frac{p(\bY \mid z_v=h, \bz_{-v})}{p(\bY_{-v} \mid \bz_{-v})},
\end{eqnarray}
where $\bY_{-v}$ is the $(V-1) \times (V-1)$ adjacency matrix without the row and column referring to node $v$. Recalling \citet{schmidt_2013}, the last term in \eqref{eq_exact_full} can be simplified as
\begin{eqnarray} \label{eq_simpl_lhd}
\frac{p(\bY \mid z_v=h, \bz_{-v})}{p(\bY_{-v} \mid \bz_{-v})}=\prod\nolimits_{k=1}^{H} \frac{\mbox{B}(a+m_{hk}^{-}+r_{vk},b+\overline{m}_{hk}^{-}+\overline{r}_{vk})}{\mbox{B}(a+m_{hk}^{-},b+\overline{m}_{hk}^{-})},
\end{eqnarray}
\noindent where $ m_{hk}^{-} $ and $ \overline{m}_{hk}^{-} $ denote the number of edges and non-edges between clusters $ h $ and $ k $,  without counting node $ v $, while $ r_{vk} $ and $\overline{r}_{vk}$ define the number of edges and non-edges between node $ v $ and the nodes in cluster~$ k $. The prior term $ \mbox{pr}(z_v=h \mid \bX, \bz_{-v}) $ in~\eqref{eq_exact_full} is directly available  from either~\eqref{eq_urn1} or \eqref{eq_urn2}, depending on whether node attributes are excluded or included, respectively. In particular, the unsupervised Gibbs-type priors discussed in  Section~\ref{sec_2.2.1} yield 
\begin{eqnarray} \label{eq_urn3}
\begin{aligned}
& \mbox{pr}(z_v=h \mid \bX, \bz_{-v})=\mbox{pr}(z_v=h \mid \bz_{-v}) \propto \begin{cases} \mathcal{W}_{V,H^{-}} (n_h^{-}-\sigma) &  \text{for} \  h \leq H^{-}, \\ \mathcal{W}_{V,H^{-}+1} &  \text{for} \  h=H^{-} + 1, \\ \end{cases} 
\end{aligned}
\end{eqnarray}
where $ n_h^{-} $ and $H^{-}$ are the cardinality of cluster $ h $ and  the total number of occupied clusters, respectively, after removing node~$ v $. Whereas, the supervised extension  in Section~\ref{sec_2.2.2} leads to
\begin{eqnarray} \label{eq_urn4}
\begin{aligned}
& \mbox{pr}(z_v=h \mid \bX, \bz_{-v}) \propto \begin{cases} \frac{n_{hx_v}^{-}+\alpha_{x_v}}{n_h^{-}+\alpha_0}\mathcal{W}_{V,H^{-}} (n_h^{-}-\sigma) &  \text{for} \  h \leq H^{-}, \\ \frac{\alpha_{x_v}}{\alpha_0}\mathcal{W}_{V,H^{-}+1} &  \text{for} \  h=H^{-} + 1, \\ \end{cases} 
\end{aligned}
\end{eqnarray}
where $ n_{h x_v}^{-} $ is the number of nodes in cluster $ h $ with covariate value $ c=x_v $,  without counting node $ v $, whereas $\alpha_{x_{v}}$ is the parameter for the category $c=x_{v}$ of node $v$. Under the priors in Table \ref{tab_1}, both \eqref{eq_urn3} and  \eqref{eq_urn4} admit the simple  expressions reported in the four Examples in Section~\ref{sec_2.2.1}.

\begin{algorithm}[t]
\vspace{3pt}
	At each iteration of the Gibbs sampler, update the cluster assignments $z_1, \ldots, z_V$ as follows:\\
	\vspace{5pt}
\textbf{For} $v=1, \ldots, V$ \textbf{do}:    
    	\vspace{3pt}
\begin{enumerate}
\item {Remove node $ v $ from the network;}
\item{If the cluster which contained node $ v $ becomes empty, discard it and relabel the group indicators  (so that clusters $1, \ldots,H^{-}$ are non-empty);}
\item{Sample $z_{v}$ from the categorical variable with probabilities as in \eqref{eq_exact_full} for $h=1,\ldots,H^{-}{+}1$, where ${p(\bY \mid z_v=h,\bz_{-v})}/{p(\bY_{-v} \mid \bz_{-v})}$ is defined in~\eqref{eq_simpl_lhd}, whereas $ \mbox{pr}(z_v = h \mid \bX, \bz_{-v}) $ coincides  with either~\eqref{eq_urn3} or  \eqref{eq_urn4}  depending on whether node attributes are excluded or included, respectively.}
\end{enumerate}
	\caption{\footnotesize{\bf Gibbs sampler for \textsc{ESBM}}\label{alg_esbm}}
\end{algorithm}

Although Algorithm~\ref{alg_esbm} leverages likelihood~\eqref{eq_marg_lhd} with the block probabilities $ \theta_{hk} $   integrated out, a plug-in estimate for each $ \theta_{hk} $ can be easily obtained. In particular, since $(\theta_{hk} \mid \bY, \bz) \sim \mbox{Beta}(a+m_{hk}, b+\overline{m}_{hk}) $, a reasonable point estimate for $ \theta_{hk} $ is
\begin{eqnarray}
\widehat{\theta}_{hk}=\mathbb{E}({\theta_{hk} \mid \bY, \bz = \widehat{\bz})= \frac{a+\widehat{m}_{hk}}{a+\widehat{m}_{hk}+b+\widehat{\overline{m}}_{hk}}}, 
\label{eq5}
\end{eqnarray}
for every $h=1, \ldots, \hat{H}$ and $k=1, \ldots, h$, where $\widehat{m}_{hk}$ and $\widehat{\overline{m}}_{hk}$ denote the number of edges and non-edges between nodes in groups $h$ and $k$, computed from the estimated  $\widehat{\bz}$.  In the next subsection, we describe improved methods for estimation of $\bz$, uncertainty quantification in group detection, model selection, and prediction.  


\subsection{Estimation, uncertainty quantification, model selection, prediction}
\label{sec_3.2}
While algorithmic methods return a single estimated partition, \textsc{esbm} provides the whole posterior distribution over the space of node partitions. To fully exploit this posterior and perform  inference directly on the space of partitions, we adapt the decision-theoretic approach of \cite{wade2018} to the block modeling setting. In this way, we summarize posterior distributions on partitions leveraging the \emph{variation of information}~(\textsc{vi}) metric \citep{meilua2007comparing}, that quantifies distances between two clusterings by comparing their individual and joint entropies, and ranges from 0 to $ \log_2 V $. Intuitively, \textsc{vi} measures the amount of information  in two clusterings relative to the information shared between them, thus providing a metric that decreases to 0 as the overlap between two partitions grows; see  \cite{wade2018} for a discussion of the key properties of \textsc{vi}. Under this framework, a  formal Bayesian point estimate for $\bz$ is that partition with the  lowest posterior averaged \textsc{vi} distance from the other clusterings, thus obtaining
\begin{eqnarray}\label{eq_estimate}
\hat{\bz}= {\arg\min}_{\bz'}\: \mathbb{E}_{\bz}[\textsc{vi}(\bz,\bz') \mid \bY],
\end{eqnarray}
where the expected value is taken with respect to the posterior  of $\bz$. Due to the huge cardinality of the space of partitions, even for moderate $V$, the optimization in~\eqref{eq_estimate} is typically carried out through a greedy algorithm \citep{wade2018}, as in the \texttt{R} package \texttt{mcclust.ext}.

The \textsc{vi} distance also provides natural strategies to construct credible sets around point estimates.  In particular, one can define a $1-\alpha$ credible ball around $ \hat{\bz} $ by ordering the partitions according to their \textsc{vi} distance from $ \hat{\bz} $, and defining the ball as containing all the partitions having less than a threshold distance from $ \hat{\bz} $, with this threshold chosen to minimize the size of the ball while ensuring it contains at least $1-\alpha$ posterior probability.   Summarizing this ball is non-trivial given the high-dimensional discrete nature of the space of partitions. In practice, as illustrated in our studies, one can report the partition at the edge of the ball, which we call  credible bound.  This form of uncertainty quantification complements the commonly-studied \emph{posterior similarity matrix} that measures, for each pair of nodes, the relative frequency of \textsc{mcmc} samples in which such nodes  are assigned to the same group \citep[e.g.,][]{wade2018}. Relative to this quantity, the additional inference methods we propose are conceptually and practically more appealing as they allow estimation and uncertainty quantification directly on the space of partitions. 

Another key inference task is selection among several candidate models --- that mainly arise in our context from the choice among different priors for $\bz$  in Section~\ref{sec_2.2}. One possibility to formally  address this goal is through the  Bayes factor \citep[e.g.,][]{kass1995}. However, this strategy requires  calculation of the marginal likelihood $p(\bY\mid \mathcal{M})=\sum_{\bz} p(\bY \mid \bz) p(\bz \mid \mathcal{M})$ for a generic model $ \mathcal{M} $, which is not available analytically under the priors in Section~\ref{sec_2.2}. Although simple strategies, such as the harmonic mean estimate  \citep{raftery2006}, can be employed to compute $p(\bY\mid \mathcal{M})$ in \textsc{sbm}s \citep[e.g.,][]{legramanti2020test}, these solutions may face instabilities and slow convergence in general settings \citep[e.g.,][]{lenk2009,pajor2017,wang2018}. To overcome these shortcomings and provide a general-use model selection strategy, we opt for the \textsc{waic} information criterion  \citep{watanabe2010,watanabe2013,gelman2014}. Relative to other  information criteria commonly employed also in the \textsc{sbm} framework and its extensions \citep[e.g.,][]{gormley2010,come2015,saldana2017,rastelli2018,lee2019}, the  \textsc{waic} yields practical and theoretical advantages \citep{gelman2014}, and has direct connections with Bayesian  leave-one-out cross-validation \citep{watanabe2010}, thus providing also a measure of edge predictive accuracy. In addition, calculation of the \textsc{waic} only requires posterior samples of the log-likelihoods for the edges $\log p(y_{vu}\mid \bz, \bTheta)=y_{vu}\log\theta_{z_v,z_u}+(1-y_{vu})\log(1-\theta_{z_v,z_u})$, $v=2, \ldots, V$, $u=1, \ldots, v-1$. These quantities can be readily obtained by combining the posterior samples for $\bz$ from Algorithm~\ref{alg_esbm}, with those for the block probabilities in $\bTheta$, which can be easily simulated from the conjugate full conditional distributions  $(\theta_{hk} \mid \bY,\bz) \sim \mbox{Beta}(a+m_{hk}, b+\overline{m}_{hk})$ for $h=1, \ldots, H$ and $k=1, \ldots, h$ via a separate algorithm that can be run in parallel across blocks and samples; see Section 3.4 in \citet{gelman2014} for details on the \textsc{waic}, and refer to the  \texttt{WAIC} function in the \texttt{R} package  \texttt{LaplacesDemon} for practical implementation. As a global measure of goodness-of-fit we also study the misclassification error when predicting each $y_{vu}$ with $\hat{\theta}_{\hat{z}_v\hat{z}_u}$ from~\eqref{eq5}.

Recalling the criminal network application in Section~\ref{sec_1.1}, predicting the group membership $z_{V+1}$ for a newly observed suspect $V+1$ is also of fundamental interest in these contexts. While common algorithmic strategies would require heuristic procedures, the urn scheme representation \eqref{eq_urn1}  of the Gibbs-type priors  provides a natural construction to obtain formal estimates of group probabilities for incoming suspects, without conditioning on external attributes that are typically unavailable in early investigations of such new individuals. Combining equations \eqref{eq_exact_full}--\eqref{eq_simpl_lhd} with the urn scheme in  \eqref{eq_urn1}, a plug-in estimate for the predictive probabilities of the cluster allocations for node $V+1$ is
\begin{eqnarray}\label{predict}
\begin{split}
&\mbox{pr}(z_{V+1}= h \mid \bY, \by_{V+1}, \hat{\bz}) \\
&\propto \mbox{pr}(z_{V+1}= h \mid  \hat{\bz})\prod\nolimits_{k=1}^{\hat{H}} \frac{\mbox{B}(a+\widehat{m}_{hk}+\widehat{r}_{V+1,k},b+\widehat{\overline{m}}_{hk}+\widehat{\overline{r}}_{V+1,k})}{\mbox{B}(a+\widehat{m}_{hk},b+\widehat{\overline{m}}_{hk})},
\end{split}
\end{eqnarray}
for each $h=1,\ldots, \hat{H}+1$, with $\mbox{pr}(z_{V+1}=h \mid \hat{\bz})$ as in  \eqref{eq_urn1}.  In  \eqref{predict},  $\by_{V+1}=(y_{V+1,1}, \ldots, y_{V+1,V})^{\intercal}$ is the vector of  newly observed edges between node $V+1$ and those already in network $\bY$. The frequencies $\widehat{m}_{hk}$ and $\widehat{\overline{m}}_{hk}$ denote instead the number of edges and non-edges between the existing nodes in groups $h$ and $k$ computed from the estimated cluster assignments in $\widehat{\bz}$, whereas $\widehat{r}_{V+1,k}$ and $\widehat{\overline{r}}_{V+1,k}$ define the number of edges and non-edges between the incoming node $ V+1 $ and the existing nodes in cluster~$ k $, still evaluated at the estimated partition $\widehat{\bz}$. Note that, under the priors in Table \ref{tab_1}, the quantity $\mbox{pr}(z_{V+1}=h \mid \hat{\bz})$ admits the closed-form expressions reported in the four Examples of Gibbs-type priors in Section~\ref{sec_2.2.1}.


\section{Simulation Studies}\label{sec_4}

\begin{figure}[b]
	\centering
\includegraphics[width=16cm]{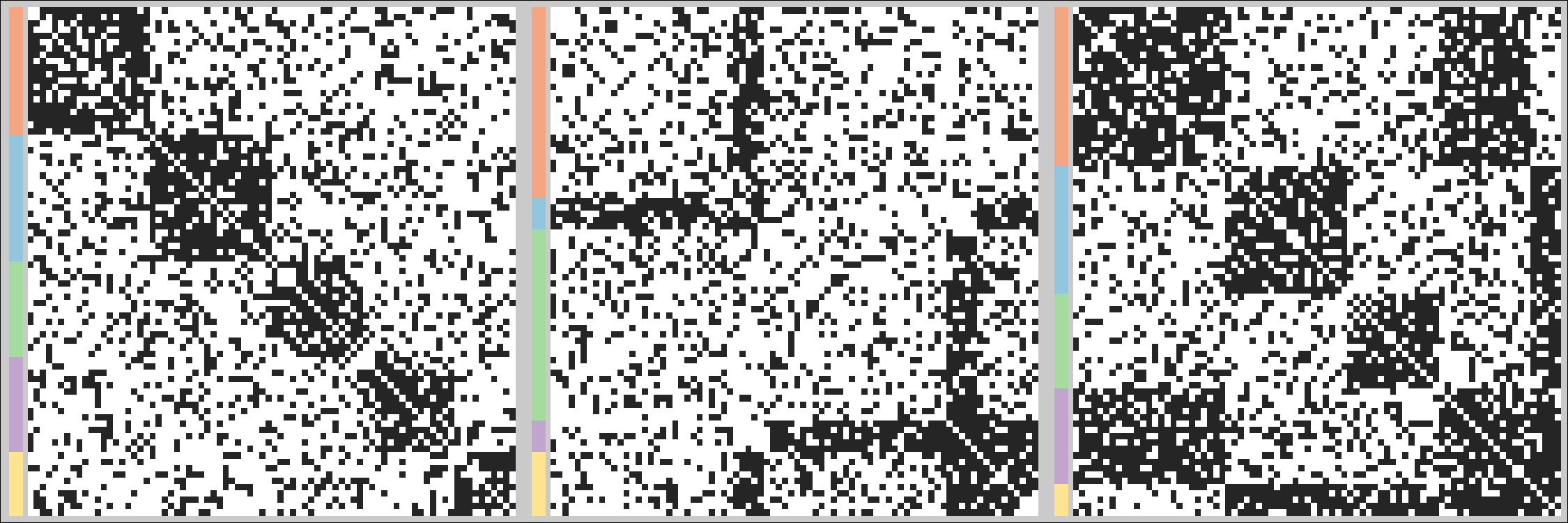}
	\put(-80,-15){(c)}
		\put(-230,-15){(b)}
		\put(-385,-15){(a)}
	\caption{Simulated adjacency matrices under the first (a), second (b) and third (c) scenario. Side colors correspond to the true partition $\bz_0$. Black cells refer to edges, whereas white cells denote non-edges. }
	\label{fsimu1}
\end{figure}

To assess the performance of \textsc{esbm}  in settings mimicking our motivating application, and quantify the advantages over state-of-the-art alternatives \citep{von2007,blondel_2008,amini2013,zhang2016,come2021}, we consider three simulated networks of $ {V=80} $ nodes displaying different criminal block structures sampled from a \textsc{sbm} with $ {{H}_0=5} $ groups, and block probabilities equal to either $ 0.75 $ or $0.25$. As shown in Figure~\ref{fsimu1}, the first network defines a horizontal criminal organization characterized by classical community structures of varying size. The second network provides, instead, a more challenging  scenario which exhibits a nested hierarchy of core-periphery, weak-community and disassortative patterns characterizing a vertical criminal organization. In particular, we assume the presence of two equally-sized macro-groups, each having a small fraction of bosses that interact with all the affiliates of the associated group and with an additional cluster of higher-level bosses. Finally, the last simulated network resembles more closely the block structures of the {\em Infinito network}, where we expect community patterns among the affiliates in each {\em locale}, core-periphery structures between such affiliates and the corresponding bosses, and assortative behaviors among the   bosses of the different {\em locali}, resulting from coverting strategies.

As we will illustrate in Table~\ref{tab_3}, state-of-the-art strategies \citep{von2007,blondel_2008,amini2013,zhang2016,come2021} applied to these three networks mostly fail in recovering the true underlying blocks and show a general tendency to over-collapse different groups, possibly due to their inability to incorporate unbalanced noisy partitions and effectively exploit attribute information. Such results motivate implementation of  \textsc{esbm}, both without and with node attributes coinciding, in this case, with the true partition $\bz_0$. This choice is useful for assessing to what extent the  supervised Gibbs-type priors and relevant competitors can effectively exploit truly informative node attributes.

\begin{table}[t]
	\centering
	\caption{{Performance of \textsc{esbm} in the three scenarios with $ H_0 = 5 $, when excluding attributes  (\textsc{unsup}), and when supervising each prior with the true partition $\bz_0$ as attribute (\textsc{sup}). Performance is measured by the \textsc{waic}, the posterior mean $\mathbb{E}[\textsc{vi}(\bz,\bz_0) \mid \bY]$ of the \textsc{vi} distance  from $ \bz_0 $, the posterior median number of  non-empty clusters $H$ (first and third quartiles in brackets), and the distance $\textsc{vi}(\hat{\bz},\bz_b)$ among the estimated partition $\hat{\bz}$ and the $95\%$ credible bound $\bz_b$. Bolded values denote the best performances among \textsc{unsup} priors within each column. Bolded gray cells denote the best overall performance in each column. 
		 }}
	\scalebox{0.92}{	
		\begin{tabular}{l@{\hspace*{2mm}}c@{\hspace*{1.5mm}}c@{\hspace*{1.5mm}}c@{\hspace*{5mm}}c@{\hspace*{2mm}}c@{\hspace*{2mm}}c@{\hspace*{5mm}}c@{\hspace*{1mm}}c@{\hspace*{1mm}}c@{\hspace*{5mm}}c@{\hspace*{3mm}}c@{\hspace*{3mm}}c}
			\toprule
			\vspace{-15pt}
			\\
			& \multicolumn{3}{c}{\textsc{waic}} &  \multicolumn{3}{c}{$\mathbb{E}[\textsc{vi}(\bz,\bz_0) \mid \bY]$} &  \multicolumn{3}{c}{$H$} &   \multicolumn{3}{c}{$\textsc{vi}(\hat{\bz},\bz_b)$}  \\ 
			\midrule
			\textsc{Scenario}	 &1  &2  & 3  &1 & 2&3  &1  &2&3&1&2&3 \\
			\cmidrule{1-13}
			\textsc{[unsup] dm} & $3551.0$&$3559.8$& $3540.3$ &$0.420$& $0.746$ &$0.517$& $8$ {[7,8]}  &  $6$ {[5,7]} &  $6$ {[5,6]} &$0.702$ & $0.971$ &$0.691$ \\ 
			\textsc{[unsup] dp}  &$3550.7$&$3559.5$&$3540.4$&$0.414$& $0.736$&$0.514$ &$7$ {[7,8]}  & $6$ {[5,7]} & $6$ {[5,6]} & $0.694$&$0.955$ &$0.692$\\ 
			\textsc{[unsup] py} & $3551.4$&$3559.0$& $3540.2$&$0.376$&$0.708$&$0.498$  & $7$ {[6,9]} &  $6$ {[5,7]} & $6$ {[5,6]} & $0.696$ &$0.884$ & $0.645$ \\ 
			\textsc{[unsup] gn} & ${\bf 3550.1}$&${\bf 3554.3}$&${\bf 3535.9}$&${\bf 0.292}$&${\bf 0.642}$& ${\bf 0.455}$&${\bf 5}$ {\bf {[5,6]}} & \cellcolor{black!10}   ${\bf 5}$ {\bf {[5,5]}} \ &\cellcolor{black!10} ${\bf 5}$ {\bf {[5,5]}} \ & $0.592$&  $0.827$&$0.601$\\ 
			\cmidrule{1-13}
			\textsc{[sup] dm} &$3522.7$&$3512.6$&$3516.6$&$0.090$&$0.155$&$0.134$&$6$ {[5,6]}&$5$ {[5,6]}& \cellcolor{black!10}  ${\bf{5}}$ {\bf{{[5,5]} } }& $0.254$&  $0.316$&$0.329$\\
			\textsc{[sup] dp} &$3522.6$&$3512.5$&$3516.5$&$0.086$&$0.155$&$0.135$&$6$ {[5,6]}&$5$ {[5,6]}& \cellcolor{black!10} ${\bf{5}}$ {\bf{{[5,5]} } }& $0.249$&  $0.316$&$0.329$\\
			\textsc{[sup] py} &$3522.2$&$3511.9$&$3516.5$&$0.074$&$0.151$&$0.134$&$6$ {[5,6]}&\cellcolor{black!10}  ${\bf{5}}$ {\bf{{[5,5]} } }&\cellcolor{black!10}  ${\bf{5}}$ {\bf{{[5,5]} } }& $0.204$&  $0.316$&$0.311$\\
			\textsc{[sup] gn} & \cellcolor{black!10} ${\bf{3521.3}}$  \ &\cellcolor{black!10}  ${\bf{3510.4}}$ \ & \cellcolor{black!10}  ${\bf{3515.2}}$  & \cellcolor{black!10}  ${\bf{0.041}}$ &\cellcolor{black!10}  ${\bf{0.139}}$ &\cellcolor{black!10}  ${\bf{0.122}}$ &\cellcolor{black!10}  ${\bf{5}}$ {\bf{{[5,5]} } }&\cellcolor{black!10} ${\bf{5}}$ {\bf{{[5,5]}} } &\cellcolor{black!10}  ${\bf{5}}$ {\bf{{[5,5]}} }  & $0.139$&  $0.297$&$0.284$\\ 
			\bottomrule		
	\end{tabular}}
    \label{tab_2}
\end{table}

Within the Gibbs-type class, we first assess the four representative unsupervised priors  for $ \bz $ presented in Table~\ref{tab_1}, and then check whether introducing informative node attributes further improves the performance  in each scenario. The hyperparameters are specified so that the prior expected number $\mathbb{E}_\textsc{dm}(H)$, $\mathbb{E}_\textsc{dp}(H)$, $\mathbb{E}_\textsc{py}(H)$ and $\mathbb{E}_\textsc{gn}(H)$ of non-empty groups under the different priors is close to $10>{H}_0$, whereas Algorithm~\ref{alg_esbm} is initialized with every node in a different cluster. In this way we can check robustness of the results to hyperparameter settings and to the initialization of the Gibbs sampler. Specifically, we set $\overline{H}=50$ and $\beta=3.5/50$ for the \textsc{dm}, $\alpha=3$ in the \textsc{dp}, $\sigma=0.6$ and  $\alpha=-0.3$ under the \textsc{py}, and $\gamma=0.45$ for the \textsc{gn}. In implementing such models we consider the default uniform setting $a=b=1$ for the prior on the block probabilities \citep[e.g.,][]{Nowicki_2001,geng_2019}, and let $\alpha_1=\cdots=\alpha_C=1$ in \eqref{eq_g}, when including node attributes. 

From Algorithm~\ref{alg_esbm} we obtain $40000$ samples for~$ \bz $, after a conservative burn-in of $10000$.  In our experiments, inference has proven robust to different initializations of $ \bz $ in Algorithm~\ref{alg_esbm}, including extreme settings with all nodes in a single group. Nonetheless,  starting with one cluster for every node provides the best overall mixing, when monitored on the chain for the likelihood in \eqref{eq_marg_lhd} evaluated at the \textsc{mcmc} samples of $\bz$. Graphical analysis of the traceplots for such a chain suggests rapid convergence and effective mixing under all models. Algorithm~\ref{alg_esbm} provides 150 samples of $\bz$ per second when executed on an iMac with 1 Intel Core~i5 3.4 \textsc{ghz} processor and 8 \textsc{gb} \textsc{ram}, thus showing good efficiency. Table~\ref{tab_2} summarizes the performance of the four priors. 

\begin{figure}[t]
	\centering
\includegraphics[width=16cm]{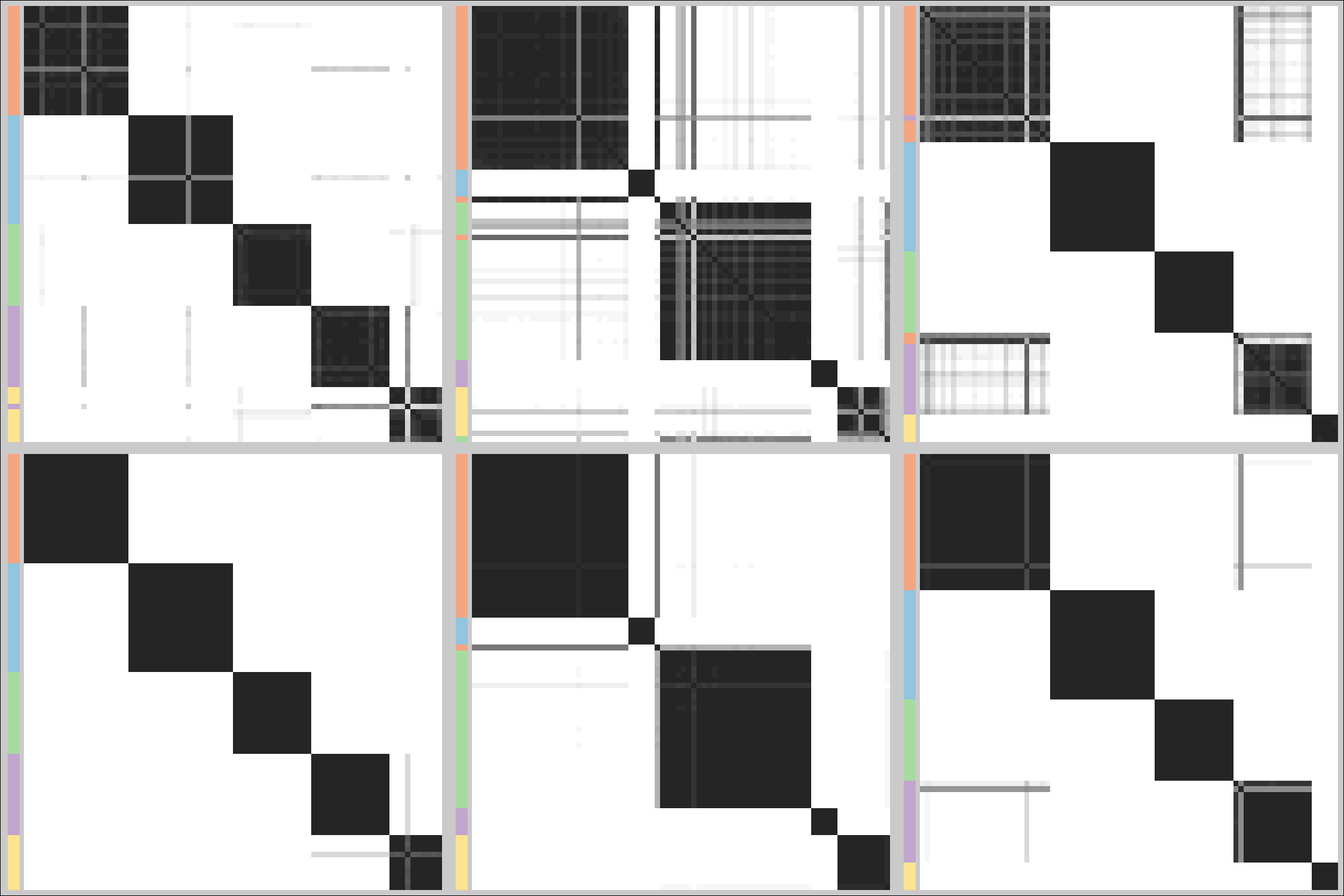}
	\put(-80,-15){(c)}
		\put(-230,-15){(b)}
		\put(-385,-15){(a)}
	\caption{For the first (a), second (b) and third (c) scenario, posterior similarity matrices under the Gnedin process from the \textsc{esbm} without (first row) and with (second row) node attributes, respectively. Cell colors range from white to black as the estimated co-clustering probability of the associated pair of nodes goes from $0$ to $1$. Side colors correspond to estimated partitions $\hat{\bz}$. }
	\label{fsimu2}
\end{figure}

Among the unsupervised Gibbs-type priors considered for $ \bz $, the Gnedin process always yields slightly improved performance in terms of \textsc{waic} and posterior mean of the \textsc{vi} distance from the true partition $ \bz_0 $. In addition, it  offers more accurate learning of the number of groups, with tighter interquartile ranges that always include the true $ {H}_0=5 $, and tighter credible balls around the \textsc{vi}-optimal posterior point estimate $ \hat{\bz} $. In our experiments, the \textsc{gn} prior was also the less sensitive to hyperparameter settings, although comparable robustness was observed even for \textsc{dm}, \textsc{dp} and \textsc{py} under moderate changes of the hyperparameters. For instance, setting these hyperparameters to induce an expected value on $H$ under all priors of $5=H_0$ instead of $10$, did not change the final conclusions provided by  Table~\ref{tab_2}.

\begin{table}[b]
			\centering
			\caption{For the three simulation scenarios, performance comparison between \textsc{esbm} with \textsc{gn} prior, and state-of-the-art unsupervised and supervised competitors in the \texttt{R} libraries \texttt{igraph}, \texttt{randnet}, \texttt{greed} and \texttt{JCDC}. These include the \texttt{Louvain} algorithm \citep{blondel_2008}, \texttt{Spectral}  clustering \citep{von2007}, \texttt{Regularized Spectral} clustering \citep{amini2013}, the \texttt{greed} clustering algorithm for \textsc{sbm} and degree corrected \textsc{sbm} (\textsc{dc}--\textsc{sbm}) \citep{come2021}, and the attribute-assisted \texttt{JCDC} community detection algorithm \citep{zhang2016}. The assessment focuses on the estimated number $\hat{H}$ of non-empty groups, the \textsc{vi} distance $\textsc{vi}(\hat{\bz},\bz_0)$ between the estimated and true partitions, and the absolute error between the estimated and true edge probabilities, averaged across the $V(V-1)/2$ node pairs.  Bolded values denote the best performances among unsupervised methods within each column, whereas bolded gray cells denote the best overall performance within each column.}
\scalebox{0.92}{
			\begin{tabular}{l@{\hspace*{12mm}}c@{\hspace*{9mm}}c@{\hspace*{9mm}}c@{\hspace*{12mm}}ccc@{\hspace*{12mm}}ccc}
			\toprule
			\vspace{-15pt}
			\\
			& \multicolumn{3}{c}{$\hat{H}$} &\multicolumn{3}{c}{$\textsc{vi}(\hat{\bz},\bz_0)$} & \multicolumn{3}{c}{\textsc{error [est]}}  \\
			\cmidrule{1-10}
			\textsc{Scenario}	& 1&2&3 &1&2&3 &1&2&3  \\
			\cmidrule{1-10}
			\textsc{[unsup] esbm (gn)} &\cellcolor{black!10}  ${\bf{5}}$ &\cellcolor{black!10}  ${\bf{5}}$ &\cellcolor{black!10}  ${\bf{5}}$ & ${\bf 0.126}$ & $0.404$ & ${\bf 0.374}$ & ${\bf  0.030}$ & $0.028$ &  $0.031$ \\ 
			 \textsc{[unsup]} \texttt{Louvain}& $4$ & $4$ & $3$ & $0.303$ & $2.904$ & $0.810$ & $0.040$ & $0.124$ & $0.051$ \\ 
			 \textsc{[unsup]} \texttt{Spectral}& $4$ & $4$ & $3$ & $0.557$ & $2.806$  & $0.810$ &$0.045$&$0.132$&$0.051$\\ 	
			 \textsc{[unsup]} \texttt{Reg.\ Spectral}&$4$ &$4$&$3$ &$0.557$&$2.634$ &$0.810$&$0.045$&$0.121$&$0.051$\\ 
			 \textsc{[unsup]} \texttt{greed} (\textsc{sbm})&$4$ &\cellcolor{black!10}  ${\bf{5}}$& $4$&$0.412$& ${\bf{0.267}}$ &$0.477$&$0.044$& ${\bf{0.027}}$& ${\bf{0.028}}$\\ 
			 \textsc{[unsup]} \texttt{greed} (\textsc{dc}--\textsc{sbm}) &$2$ &$1$& $2$&$1.469$&$1.936$ &$1.180$&$0.105$&$0.126$&$0.084$\\ 
			\cmidrule{1-10}
\textsc{[sup] esbm (gn)} &\cellcolor{black!10}  ${\bf{5}}$& \cellcolor{black!10}  ${\bf{5}}$&\cellcolor{black!10}  ${\bf{5}}$ &\cellcolor{black!10}  ${\bf{0.000}}$&\cellcolor{black!10}  ${\bf{0.159}}$&\cellcolor{black!10}  ${\bf{0.000}}$&\cellcolor{black!10} ${\bf{0.022}}$&\cellcolor{black!10}  ${\bf{0.026}}$&\cellcolor{black!10}  ${\bf{0.023}}$\\ 			
		\textsc{[sup]}	\texttt{JCDC} ($ w_n=5 $) & $4$& $4$& $3$& $0.417$& $2.825$& $0.810$& $0.040$&$0.116$ &$0.051$ \\ 	
		\textsc{[sup]}	\texttt{JCDC} ($ w_n=1.5 $)& $4$& $4$& $3$& $0.303$& $2.024$&$0.703$ &$0.040$ &$0.112$ &$0.047$ \\
			\bottomrule		
	\end{tabular}}
   \label{tab_3}
\end{table}

As expected, including informative attributes further improves performance of all unsupervised priors in each scenario, effectively lowering $ \mathbb{E}[\textsc{vi}(\bz,\bz_0) \mid \bY] $, and further shrinking the credible balls. In a sense, this is the best  setting, since we consider the true $ \bz_0 $ as node attribute. We also tried  supervising with a random permutation of $ \bz_0 $. This resulted in a slight performance deterioration relative to the unsupervised \textsc{gn} prior, which is doubly reassuring. In fact, on one hand it shows that, under the proposed model selection criteria, an unsupervised prior would be preferred to one with non-informative attributes. On the other, the fact that performance deterioration is not dramatic suggests  robustness in learning. According to the {\em posterior similarity matrices} in  Figure~\ref{fsimu2}, unbalanced partitions are harder to infer, especially without attributes.  However this gap vanishes when including informative attributes that  can successfully support inference and reduce posterior uncertainty.  All misclassification errors for in-sample edge prediction are about $0.24$, almost matching the one expected under the true model. This suggests accurate calibration and a tendency to avoid overfitting in \textsc{esbm}s. Such a property is further confirmed by the performance in predicting, via \eqref{predict}, the group membership for $300$ new nodes, among which $50$ are simulated from a cluster not yet observed in the original networks. For this task, the missclassification errors under the supervised \textsc{gn} prior are $0.01$, $0.08$ and $0.04$ in the first, second and third scenario, respectively.

To further clarify the magnitude of the improvements provided by the \textsc{esbm}, Table~\ref{tab_3} compares the performance of  \textsc{gn}  prior --- which proved the more accurate in  Table~\ref{tab_2} --- with the results obtained under the state-of-the-art alternatives \citep{von2007,blondel_2008,amini2013,zhang2016,come2021}  discussed in Section~\ref{sec_1.2}. Since most of these competitors are non-Bayesian and only provide a point estimate $\hat{\bz}$ of $\bz$, Table~\ref{tab_3} focuses on measures of accuracy in point estimation to facilitate comparison among the different methods. In estimating $H$ under spectral clustering, we consider a variety of model selection criteria available in the \texttt{R} library \texttt{randnet}, and set $\hat{H}$ equal to the median of the values of $H$ estimated under the different strategies. These include the Beth-Hessian solution from \citet{le2015}, the likelihood ratio strategy by \citet{wang2017}, and the cross-validation methods developed in \citet{chen2018} and \citet{li2020network}. This estimate for $H$ is also used as a sensible starting value to initialize the greedy clustering algorithm for \textsc{sbm} and \textsc{dc}--\textsc{sbm} in the \texttt{R} library \texttt{greed} \citep{come2021}. As shown in the \texttt{R} manual of the   \texttt{greed} library, this strategy estimates $\bz$ under a Dirichlet-multinomial prior for the group membership indicators. Hence, to make results comparable with the proposed \textsc{esbm} class, we set the Dirichlet hyperparameter in \texttt{greed} equal to $3.5/50$, as done for the \textsc{dm} prior under  \textsc{esbm}.  Among the available methods that leverage attribute information, we consider the community detection algorithm proposed by \citet{zhang2016}, under different default values for the tuning parameters and setting, again, $H=\hat{H}$. This strategy has been shown in \citet{zhang2016} to yield improved empirical performance relative to other powerful attribute-assisted solutions, thereby providing a suitable benchmark competitor.

As illustrated in Table~\ref{tab_3}, the above competitors display a tendency to systematically under-estimate the true number of non-empty groups, and exhibit reduced accuracy in learning the true partition and the exact edge probabilities, relative to \textsc{esbm} with  \textsc{gn}  prior. This accuracy reduction is further affected by the difficulties in learning more complex  block structures beyond communities, which affect performance even when supervising the algorithms with the true underlying partition $\bz_0$. The \texttt{greed} clustering algorithm for \textsc{sbm} \citep{come2021} is, overall, the closest in performance to the proposed \textsc{esbm} with  \textsc{gn}  prior and, in additional studies, we found that its performance can be typically improved by setting hyperparameters and starting values more extreme than those  underlying the true data generative process. While this choice is possible, in practice the truth is unknown and, hence, a more data-driven strategy to set these quantities, as the one we consider for the \texttt{greed} algorithm evaluated in Table~\ref{tab_3}, is more desirable in general. The unsupervised and supervised \textsc{esbm} with  \textsc{gn}  prior always yield accurate point estimates of $\bz$ in all scenarios and, unlike  the competitors under analysis, further allow principled uncertainty quantification and not just point estimation. As expected, the output of the greedy clustering algorithm by \citet{come2021}  in Table~\ref{tab_3} points clearly toward \textsc{sbm} rather than  \textsc{dc}--\textsc{sbm} in all the three scenarios. This result is further confirmed by state-of-the-art model selection strategies implemented in the functions \texttt{NCV.select} \citep{chen2018} and \texttt{ECV.block} \citep{li2020network} of the \texttt{R} library \texttt{randnet}. 


\section{Application to the {\em Infinito network}}\label{sec_5}

We apply the approach developed in Sections~\ref{sec_2}--\ref{sec_3} to the {\em Infinito network} presented  in Section~\ref{sec_1.1}. Despite its potential in unveiling the internal organization of 'Ndrangheta, such a network has received little attention within the statistical literature, apart from initial analyses in \citet{calderoni_2014} and \citet{calderoni2017}. These two contributions have the merit of providing early results on the relevance of block structures as key sources of knowledge to shed light on the internal architecture of criminal organizations. However, the overarching focus is on classical community structures and their relation with suspect attributes, such as {\em locali} affiliation and role. As clarified in  Section~\ref{sec_1} and in the simulation studies in Section~\ref{sec_4}, this approach rules out recurring block structures in criminal networks, fails to formally include error-prone attributes  in the modeling process, and lacks extensive methods for uncertainty quantification, model selection and prediction.

To address the above issues and obtain a deeper understanding of the  internal structure behind  {\em La Lombardia}, we provide an  in-depth analysis of the {\em Infinito network} under the \textsc{esbm} class. As for the simulations in Section~\ref{sec_4}, we first identify a suitable  candidate model by comparing the performance of the unsupervised and supervised priors  for $ \bz $ presented in Sections \ref{sec_2.2.1}--\ref{sec_2.2.2}, with hyperparameters inducing $ 20 $ expected clusters a priori. This value is four times the number of {\em locali} in the network, which seems reasonably conservative. In particular, we let $\overline{H}=50$ and $\beta=12/50$ for the \textsc{dm}, $\alpha=8$ in the \textsc{dp}, $\sigma=0.725$ and  $\alpha=-0.350$ for the \textsc{py}, and $\gamma=0.3$ under \textsc{gn}. Posterior inference relies again on $40000 $ \textsc{mcmc} samples produced by Algorithm~\ref{alg_esbm}, after a burn-in of $10000$. The  traceplots for the likelihood in \eqref{eq_marg_lhd} suggest adequate mixing and rapid convergence as  in the simulations, with similar running times. Also in this case, the results were overall robust to initialization and moderate changes in the hyperparameter settings.

\begin{table}[b]
	\centering
	\caption{Performance of \textsc{esbm} in the Infinito network, when excluding attributes  (\textsc{unsup}), and when supervising each prior with role-locale information (\textsc{sup}). Performance is measured by the \textsc{waic}. Bolded values denote best performance among the \textsc{unsup} priors. Bolded gray cells indicate best overall performance. We also provide the posterior median number of non-empty clusters $H$ (first and third quartiles in brackets), and the distance $\textsc{vi}(\hat{\bz},\bz_b)$ among the estimated partition $\hat{\bz}$ and the $95\%$ credible bound $\bz_b$. }
\scalebox{0.95}{
		\begin{tabular}{l@{\hspace*{20mm}}cc@{\hspace*{10mm}}cc@{\hspace*{10mm}}cc}
			\toprule
			\vspace{-15pt}
			\\
			& \multicolumn{2}{c}{\textsc{waic}} &   \multicolumn{2}{c}{$H$} &  \multicolumn{2}{c}{$\textsc{vi}(\hat{\bz},\bz_b)$}    \\ 
			\cmidrule{1-7}
			&\textsc{unsup} & \textsc{sup} &  \textsc{unsup} & \textsc{sup} &  \textsc{unsup} & \textsc{sup}   \\ 
						\cmidrule{1-7}
			\textsc{dm}& $1228.5$&$1199.0$&  $14$ \footnotesize{[14,15]}  &  $15$ \footnotesize{[15,15]} &$0.279$ &$0.163$ \\ 
			\textsc{dp}& $1256.2$&$1198.5$ & $14$ \footnotesize{[14,14]} &   $15$ \footnotesize{[15,16]}&$0.219$&$0.279$ \\ 
			\textsc{py}& $1279.9$&$1225.5$ & $14$ \footnotesize{[14,14]} &  $15$ \footnotesize{[14,15]} &$0.299$&$0.199$ \\ 
			\textsc{gn} &  ${\bf 1204.7}$& \cellcolor{black!10} ${\bf{1194.1}}$&  $15$ \footnotesize{[15,15]} &$15$ \footnotesize{[15,16]}  &$0.317$& $0.221$\\ 
					\bottomrule		
	\end{tabular}}
	    \label{tab_2_app}
\end{table}

As clarified in Table~\ref{tab_2_app}, \textsc{gn} yields the best performance also in the {\em Infinito network}, relative to the other examples of Gibbs-type priors commonly implemented in network studies. This provides quantitative support for the conjecture  in Section~\ref{sec_2.2.1} on the suitability of  \textsc{gn} as a realistic prior for grouping structures in organized crime. Moreover, as seen in Table~\ref{tab_2_app}, supervising the priors with the additional information on role and {\em locale} affiliation  leads to a further reduction in the \textsc{waic} and lower posterior uncertainty, meaning that  such attributes carry information about  'Ndrangheta  modules.  \citet{calderoni_2014} and \citet{calderoni2017} investigated similar effects, but with a focus on descriptive analyses of classical community structures, thus obtaining results that partially depart from the expected vertical  architecture of 'Ndrangheta  \citep[][]{paoli2007,catino2014}. In fact, the authors obtain communities defined by unions of multiple {\em locali}, and seem unable to separate affiliates from bosses throughout the partition process. As shown in panel (a) of Figure~\ref{F2}, this tendency is confirmed when applying the Louvain algorithm \citep{blondel_2008} to the {\em Infinito network}. Compared to the \textsc{esbm} in panel (c) of Figure~\ref{F2}, the Louvain algorithm provides an overly coarsened view of the block structures in the {\em Infinito network}.

\begin{figure}[t]
	\centering
\includegraphics[width=15.5cm]{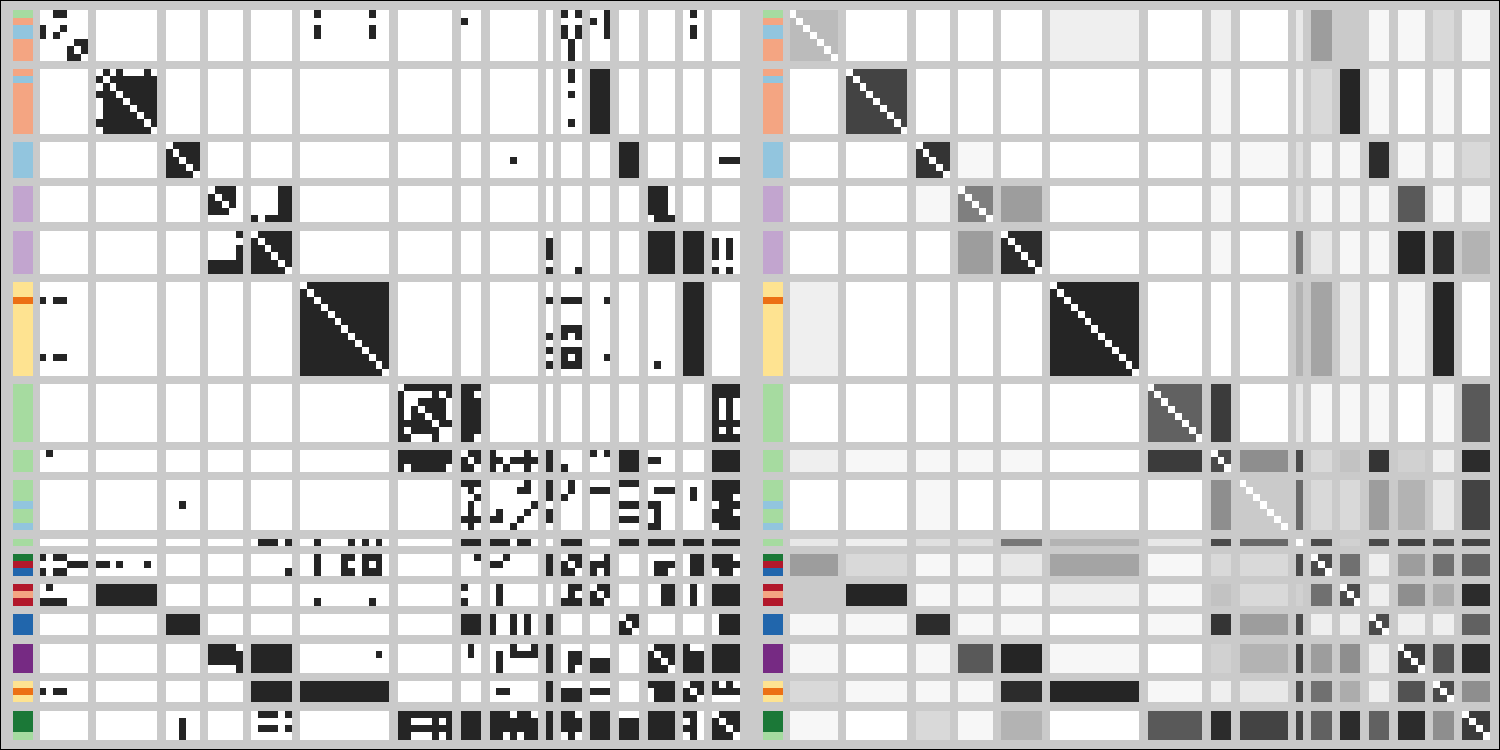}
		\put(-115,-15){(b)}
		\put(-335,-15){(a)}
	\caption{Adjacency matrix (a) and estimated edge-probability matrix (b)  of the  Infinito network with nodes re-ordered and partitioned in blocks according to the clustering structure estimated under \textsc{esbm} with supervised \textsc{gn} process prior.  Side colors correspond to the different locali, with darker and lighter shades denoting  bosses and affiliates, respectively. }
	\label{F_app1}
\end{figure}

\begin{figure}[t]
	\centering
	\includegraphics[trim=1cm 0.8cm 0cm 0.5cm,clip,width=12.9cm]{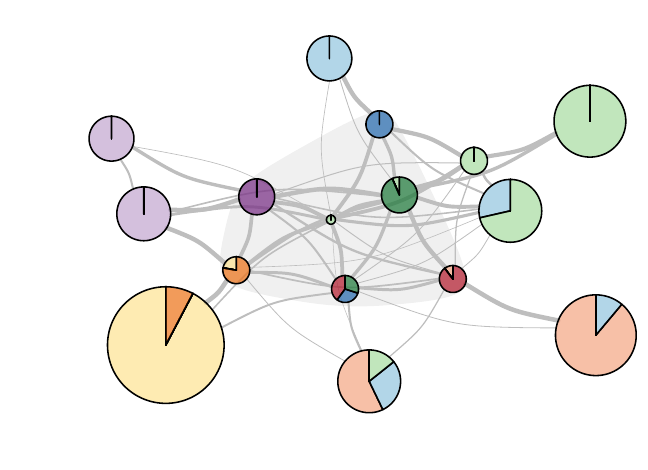}
	\caption{Network representation of the inferred clusters in the Infinito network. Each node denotes one cluster and edges are weighted by the estimated block probabilities. Node sizes are proportional to cluster cardinalities, while pie-charts represent compositions with respect to locale affiliations and leadership role; colors are the same as in Figure~\ref{F_app1}. To provide more direct insights, the composition with respect to role in the smaller-sized pie-charts is  re-weighted to account for the fact  that bosses are less frequent in the network relative to affiliates.  Node positions are obtained via force-directed placement  \citep{fru1991} to reflect  strength of connections. }
	\label{F_app2}
\end{figure}

Recalling forensic theories on organized crime  \citep[e.g.,][]{paoli2007,catino2014}, our conjecture is that 'Ndrangheta displays more complex block structures in which the pure communities among the affiliates within each {\em locale} are combined with higher-level core-periphery coordinating structures between the bosses. Unlike classical community detection algorithms, the \textsc{esbm} crucially accounts for these architectures, thus providing unprecedented empirical evidence in support of such forensic theories, as seen in Figures~\ref{F_app1}--\ref{F_app2}. These graphical assessments are based on a point estimate $\hat{\bz}$ of the partition structure under the supervised  \textsc{gn} process prior, which we consider in the subsequent analyses of the  {\em Infinito network}, due to its superior performance in Table~\ref{tab_2_app} and the relatively low posterior uncertainty around the estimated partition $\hat{\bz}$ --- the radius of the credible ball is far below the maximum achievable \textsc{vi} distance  of $ \log_2 84 \approx 6.392$.  To formally confirm the  forensic hypotheses, we compute the difference in \textsc{waic} between the unsupervised and supervised \textsc{gn}  prior, with suspects' attribute $\bX$ defining the conjectured structure. In particular, the class of each affiliate corresponds to the associated {\em locale}, whereas all the bosses share a common label indicating that such members have  a leadership role in the organization. Moreover, a subset of the affiliates of the purple {\em locale} who are known from the judicial acts\textsuperscript{\ref{note1}} to cover a peripheral role are assigned a distinct label. The resulting difference is $10.6$, which provides a strong evidence in favor of our conjecture, when compared with the thresholds suggested for related information criteria \citep[e.g.,][]{spiegelhalter2002bayesian,gelman2014}.

As shown in Figure~\ref{F2}, such fundamental structures are hidden not only to community detection algorithms  \citep{blondel_2008}, but also to spectral clustering solutions \citep{von2007} which account for more complex  block structures. This is further confirmed by the  higher  values  for the deviance $\mathcal{D}=-2\log p(\bY \mid \hat{\bz})$ under the state-of-the-art competitors discussed in Section~\ref{sec_1}, and evaluated in Section~\ref{sec_4}. More specifically, the estimated partitions under \texttt{Louvain} \citep{blondel_2008}, \texttt{Spectral} \citep{von2007}, \texttt{Reg. Spectral} \citep{amini2013}, \texttt{greed} (\textsc{sbm}) \citep{come2021},  \texttt{JCDC} ($w_n=5$) and  \texttt{JCDC} ($w_n=1.5$) \citep{zhang2016} yield deviances of $2371.0$, $2108.7$, $1954.0$, $1601.2$, $2104.6$ and $2162.6$, respectively, whereas those obtained under the unsupervised and supervised \textsc{gn} process prior are $1552.8$ and $1548.8$, respectively. As for the simulation study, the Dirichlet hyperparameter for the  \texttt{greed} algorithm is set at the same value $12/50$ considered for the \textsc{dm} prior under \textsc{esbm} in the application. Similarly, any time an estimate or a starting value of $H$ is required to implement one of the competitors, we set it  equal to the median of the values of $H$ given by different selection strategies \citep{le2015,wang2017,chen2018,li2020network}. Since this estimate of $H$ is lower than the one obtained under the \textsc{gn} prior, we also compute the deviances leveraging the same number of non-empty clusters $\hat{H}=16$ inferred by the  \textsc{gn} process, thus providing an assessment not affected by the different model complexities. This alternative implementation yields the same conclusions, thereby confirming the superior performance of the \textsc{esbm} class also in this  application. To evaluate the plausibility of the stochastic block model assumption relative to its degree-corrected version \citep{karrer2011}, we further studied the output of the \texttt{R}  functions \texttt{NCV.select} \citep{chen2018} and \texttt{ECV.block} \citep{li2020network} in the \texttt{R} library \texttt{randnet}, which allow to formally select between \textsc{sbm} and \textsc{dc}--\textsc{sbm}. Both strategies provide support in favor of \textsc{sbm} in this specific application.

The above results are also confirmed  in Figure~\ref{F2}, which clearly highlights the improved ability of the supervised \textsc{gn} process prior in learning  the block structures that characterize the {\em Infinito network}. According to Figures~\ref{F_app1} and \ref{F_app2}, such modules suggest a nested partition structure mainly defined by the two macro-blocks of affiliates and bosses, which are further partitioned in sub-groups mostly coherent with the {\em locale} affiliation. The affiliates' groups typically exhibit community patterns and connect to the hidden core mainly through the bosses  of the corresponding {\em locale}, which in turn display weak assortative structures in the higher-level coordinating architecture among bosses of different {\em locali}. 

\begin{figure}[t]
	\centering
	\includegraphics[width=13.5cm]{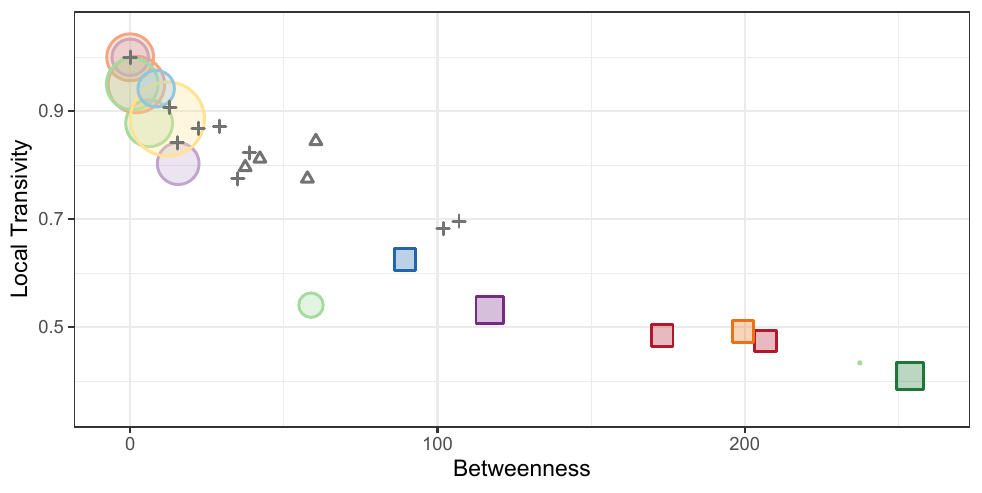}
	\caption{Scatterplot of the average betweenness and local transitivity for each estimated cluster under the supervised \textsc{gn} prior. Sizes are proportional to cluster cardinalities, whereas the color of each point is set equal to the one occupying the largest portion of the associated pie-chart in Figure~\ref{F_app2}. Circles and squares represent groups mostly referring to affiliates and bosses, respectively, while the $\Delta$ and $+$ symbols denote cluster-specific measures computed from the partitions estimated under the  Louvain algorithm \citep{blondel_2008} ($\Delta$) and spectral clustering \citep{von2007} ($+$).  }
	\label{F_app3}
\end{figure}

Figure~\ref{F_app3} confirms this result by showing how affiliates' groups are typically characterized by high local transitivity and low betweenness, whereas clusters of bosses display the opposite behavior. This is a fundamental finding  which provides new empirical evidence on the attempt of 'Ndrangheta bosses to address the tradeoff between efficiency and security \citep{morselli2007} via the creation of low-sized, sparse and secure core groups with a high betweenness that favors the flow of information towards larger and dense groups of affiliates, which guarantee efficiency. Besides these recurring architectures, the flexibility of \textsc{esbm} is also able to account  for other informative local deviations. For instance, the first group in Figure~\ref{F_app1} comprises affiliates from different {\em locali}, who were found in judicial acts\textsuperscript{\ref{note1}} to have peripheral roles. Similarly, the moderate block-connectivity patterns between the purple  {\em locale} and the  yellow one in Figures~\ref{F_app1}--\ref{F_app2}, are consistent with the fact that the latter was created as a branching of the former\textsuperscript{\ref{note1}}. The green {\em locale} has instead more complex block structures among affiliates, with a fragmentation in various subgroups denoting middle-level leadership positions. According to  the  judicial acts\textsuperscript{\ref{note1}}, these positions typically refer to authority roles in overseeing criminal actions or in guaranteeing coordination between {\em La Lombardia} and  the leading 'Ndrangheta families in Calabria. Similar roles are covered also by the small fraction of affiliates allocated to groups of bosses. Among these affiliates it is worth highlighting the suspect allocated to the single-node cluster with the most central position in Figure~\ref{F_app2}. While not being classified as a boss in the judicial acts\textsuperscript{\ref{note1}}, such a suspect is a senior member of high rank in the organization with fundamental mediating roles between all the {\em locali}, and with the leading 'Ndrangheta families in Calabria. Hence, the actual position of such an affiliate in the vertical structure of {\em La Lombardia} may be much higher than currently reported.

As shown in Figures~\ref{F2} and \ref{F_app3}, all the above structures cannot be inferred under state-of-the-art alternatives, and therefore open new avenues to obtain a substantially improved understanding of  the criminal network organization under \textsc{esbm}, along with refined predictive strategies for incoming affiliates. In particular, the predictive methods in Section~\ref{sec_3.2} applied to the $34$ held-out suspects in the {\em Infinito network} crucially allow to recognize the role of incoming criminals without the need to use external  information, that may not be available when a suspect is first observed.  In fact, classifying the  $34$ held-out suspects via \eqref{predict} with an unsupervised \textsc{gn} prior favors allocation of new affiliates to current clusters characterized by high normalized local transitivity and low normalized betweenness, whereas incoming bosses are assigned to groups with much lower difference among these quantities. More specifically, the average difference between the two measures is $0.884$ for the held-out affiliates, and $0.204$ for the held-out bosses.


\section{Discussion and future research directions}\label{sec_6}
Criminal networks provide a fundamental field of application where the advancements in network science can have a major societal impact. However, despite the relevance of such studies, there has been limited consideration of criminal networks in the statistical literature, and the focus has been largely on  restrictive methods that offer limited knowledge on the internal structure of criminal organizations. To cover this gap,  we proposed \textsc{esbm}s as a broad class of realistic models that unifies most existing \textsc{sbm}s via Gibbs-type priors. Besides providing a single methodological, theoretical and computational framework for various \textsc{sbm}s, such a generalization facilitates the proposal of new models by exploring alternative options within the Gibbs-type class, and allows natural inclusion of attributes via connections with \textsc{ppm}s. Both aspects are fundamental to investigate criminal networks. For example, we have shown in simulations that the Gnedin process, which to the best of our knowledge had never been used in \textsc{sbm}s, yields a suitable prior for partition structures in organized crime, and can improve the performance of already-implemented \textsc{dp}, \textsc{py} and \textsc{dm} in various realistic criminal networks where routine strategies, such as community detection and spectral clustering, fail. The motivating {\em Infinito network} application clarifies the benefits of our extended class of models and  methods,  providing formal unprecedented empirical evidence to several forensic theories on the internal functioning of complex criminal organizations, such as 'Ndrangheta. 

The present work offers also many future directions of research. For example, the highly general and modular structure of \textsc{esbm}s motivates application to modern real-world networks beyond criminal ones, and facilitates extensions to directed, bipartite and weighted networks. To address this goal, it is sufficient to substitute the beta-binomial likelihood in \eqref{eq_marg_lhd} with suitable ones, such as gamma-Poisson for count edges and Gaussian--Gaussian for continuous ones. Other types of suspect attributes beyond categorical ones can also be easily included leveraging the default choices suggested by \cite{muller2011product} for $ p(\cdot) $ in \eqref{eq_ppmX} under continuous, ordinal and count-type attributes. Additional applications to other criminal networks and further extensions to alternative representations, such as the mixed membership \textsc{sbm} \citep{air_2008,ranciati2020} and degree corrected \textsc{sbm} \citep{karrer2011}, are also worthy of exploration.  Despite the relevance of such constructions, we shall emphasize that  while  \textsc{esbm} preserves interpretability and parsimony by avoiding mixed membership structures, it still allows quantification of uncertainty in the degree of affiliation to different groups via formal inference on the posterior similarity matrix and on the credible bounds. Finally, although studying the coverage properties of the credible balls presented in Section~\ref{sec_3.2} is still an ongoing area of research  that goes beyond the scope of the present article \citep{wade2018}, it would be of interest to empirically check such properties within the  \textsc{esbm}  context.



\end{document}